\documentclass[3p,times]{elsarticle}
\usepackage{epsfig,graphics,amssymb,amsmath,graphicx,indentfirst,subfig,float,appendix,multirow,pifont,color,soul,esint}
\usepackage{booktabs}
\usepackage{threeparttable}
\usepackage{graphicx}
\newcommand{\bl}[1]{\boldsymbol{#1}}

\begin{document}
\begin{frontmatter}

\title{A scalable spectral Stokes solver for simulation of time-periodic flows in complex geometries}

\author[cornell]{Chenwei Meng}
\author[cornell]{Anirban Bhattacharjee}
\author[cornell]{Mahdi Esmaily}
\address[cornell]{Sibley School of Mechanical and Aerospace Engineering, Cornell University, Ithaca, NY 14850, USA}

\date{\today} 
\begin{abstract}
Simulation of unsteady creeping flow in complex geometries has traditionally required the use of a time-stepping procedure, which is typically costly and unscalable. 
To reduce the cost and allow for computations at much larger scales, we propose an alternative approach that is formulated based on the unsteady Stokes equation expressed in the spectral domain. 
This transformation results in a boundary value problem with an imaginary source term proportional to the computed mode that is discretized and solved in a complex-valued finite element solver using Bubnov-Galerkin formulation. 
This transformed spatio-spectral formulation presents several advantages over the traditional spatio-temporal techniques. 
Firstly, for cases with boundary conditions varying smoothly in time, it provides a significant saving in computational cost as it can resolve time-variation of the solution using a few modes rather than thousands of time steps. 
Secondly, in contrast to the traditional time integration scheme with a finite order of accuracy, this method exhibits a super convergence behavior versus the number of computed modes.
Thirdly, in contrast to the stabilized finite element methods for fluid, no stabilization term is employed in our formulation, producing a solution that is consistent and more accurate.
Fourthly, the proposed approach is embarrassingly parallelizable owing to the independence of the solution modes, thus enabling scalable calculations at a much larger number of processors. 
The comparison of the proposed technique against a standard stabilized finite element solver is performed using two- and three-dimensional canonical and complex geometries. 
The results show that the proposed method can produce more accurate results at 1\% to 11\% of the cost of the standard technique for the studied cases.

\end{abstract}
\end{frontmatter}




\section{Introduction} \label{sec:introduction}
Fast and cost-effective simulation of flow in complex geometries is highly desirable, at least for two sets of applications. 
The first is the time-critical problems where the results must be obtained in a given time window, thus imposing an upper bound on the time-to-solution. 
The second are those with a bound on the computational cost, where the number of CPU-hours spent on the simulation must be limited.
While these two restrictions on time and cost are identical for sequential simulations, they may well diverge for parallel implementations depending on the scalability of the application.

A prominent area where all the above conditions and restrictions apply is cardiovascular modeling. 
The flow is unsteady and time-periodic, the geometry is often complex, the cost must remain low, and the available time for simulation is often limited. 
A specific example within this realm is patients with congenital heart disease who must undergo an operation right after birth \cite{blalock1945surgical}. 
Utilizing cardiovascular modeling as a part of patient-specific surgical planning will require the simulation to be completed between diagnosis (birth) and the time of operation (a few days after birth at most) \cite{marsden2015multiscale}. 
Moreover, performing shape optimization to improve surgical anatomy relies on having a highly efficient computational framework since a formal optimization study typically requires tens to hundreds of simulations \cite{esmaily2012optimization}. 
A similar constraint exists on the efficiency of the underlying computational method if such computations were to be performed outside of the academic setting and at an industrial scale, where access to high-performance computing resources for a large number of patients remains limited due to the cost considerations \cite{pekkan2008patient}.

While there exists a broad range of numerical methods for simulating flow in complex geometries, here we consider stabilized finite element techniques to establish the relative accuracy and cost-efficiency of our method since these methods are widely adopted for cardiovascular blood flow simulation \cite{steinman2013variability} and closest to the method proposed in this study.
Within the realm of stabilized finite element methods (FEM), we particularly focus on the residual-based variational multiscale method (RBVMS), which is constructed on top of streamwise upwind Petrov-Galerkin (SUPG) technique \cite{brooks1982streamline, bazilevs2007variational, hsu2010improving}. 

As is the case with the majority of the time-dependent partial differential equations solvers, the application of the RBVMS technique to a time-periodic problem (e.g., blood flow in the circulatory system) requires discretization of the underlying governing equations in time.
Thus, the cost of these methods is inherently dependent on the time step size, increasing as time step size decreases. 
While the maximum allowable time step size is limited by the stability considerations for explicit time integration schemes, it is often more relaxed for implicit schemes, where it is determined solely based on the desired accuracy \cite{hsu2002implicit, zhang2004calculations}.
The larger time step size for implicit schemes does not necessarily imply their lower cost, as the fewer time steps can be offset by the larger number of Newton-Raphson iteration per time step.
Furthermore, the high cost associated with the time integration is exacerbated by the transient nature of the solution that requires simulation of many cycles to achieve cycle-to-cycle convergence.
In addition to these setbacks, one well-known issue associated with the RBVMS formulation is its lack of consistency with regard to time step size that, for instance, leads to the time step size dependency of the final solution obtained from a steady-state problem.

A remedy for the aforementioned issue is to transfer the temporal problem to a frequency domain using Fourier transformation and subsequently solve the transformed boundary value problem mode-by-mode.
Since the resulting spectral method bypasses the time integration, its accuracy and stability are not affected by the time integration scheme. 
We should note that the number of modes that must be incorporated into such calculation, at least in theory, must be equal to the number of time steps for all time scales in the problem to be captured. 
However, for cardiovascular flows, the majority of those modes can be neglected with minimal error, as we will show later in this article. 
In practice, a few modes (less than ten) suffices for sufficiently smooth and time-periodic boundary conditions, thus offering a significant reduction in overall cost in comparison to a traditional solver that requires thousands of time steps for simulation of a single cardiac cycle \cite{rosenfeld1995utilization}.
On top of this saving, the spectral technique can save on the computation required for cycle-to-cycle convergence as it is based on a boundary value problem that does not include a transient solution.  
Additionally, for the linear Stokes equation that is the focus of this study, all the solution modes are orthogonal and independent, which present two major advantages. 
 Firstly, capturing a new mode in the solution (either due to a modification in the boundary condition or stricter tolerance on the solution) can be done by solving one additional boundary value problem as there is no need to recompute previous modes. 
The minimal cost can be contrasted against the costly alternative of a traditional time integration scheme in which the entire computation must be repeated with a new time step size if higher accuracy is desired. 
Secondly, the spectral technique can be implemented as an embarrassingly parallelizable scheme with each group of processors computing the solution associated with only one of the many modes. 
Therefore, by formulating the original temporal problem as a set of orthogonal boundary value problems, one can significantly reduce the total computational cost and, at the same time, scale computations to a much larger number of processors.

The enumerated advantages of formulating a problem in the spectral domain have attracted researchers to investigate such schemes in the past. 
Perhaps the most relevant scheme to the present study that has come out of such efforts is the development of the Time Spectral Method, which has been primarily employed by the aerospace community for simulation of compressible flows over bluff bodies, such as blades in a compressor rotor and a pitching airfoil \cite{hall2002computation, mcmullen2001acceleration, mcmullen2003application, gopinath2005time}.
In this paper, we propose a new spatio-spectral technique named Complex-valued Stokes Solver (SCVS) that can be differentiated from the Time Spectral Method in a number of ways. 
First and foremost, the aforementioned Time Spectral Methods are formulated for the Euler's equations that are nonlinear and the SCVS for the Stokes' equations that are linear. 
This nonlinearity that creates a coupling between all modes is dealt with by establishing a relationship between solutions at all time steps and developing a pseudo-time-advancement technique that solves for all the modes (i.e., time steps) at once. 
In contrast, the SCVS exploits the linearity of the governing equations and is directly formulated in the spectral domain. 
As a result, the SCVS is implemented using complex arithmetic in contrast to the Time Spectral Method that is based on pure real arithmetic. 
Thirdly, we specifically introduce the SCVS for the simulation of creeping blood flow in the cardiovascular system. 
What distinguishes cardiovascular systems from prior applications is the geometrical complexity, as such models typically involve multiple inlets and outlets that may be coupled to lower-order models as boundary conditions \cite{moghadam2013modular}. 
Fourthly, in contrast to the aforementioned studies that employ finite volume technique for spatial discretization, we experiment with the finite element method in the SCVS as a widely adopted technique for cardiovascular modeling \cite{taylor2009patient, bazilevs2013computational}.
Lastly, in this study, we discuss in detail the convergence, accuracy, and computational efficiency of the SCVS as a function of element size, number of modeled modes, and the underlying linear solver tolerance.

The paper is organized as follows: we introduce the governing equations, namely Stokes equation with a complex-valued source term, in the spectral domain  in Section 2.1.  
Second, we present  the weak  and discrete formulations of those equations based on finite element formulation  in Sections 2.2 and 2.3, respectively.
 Then, we introduce the SCVS algorithm and elaborate on how to estimate the error in Sections 2.4 and 2.5. Next, we move on to Section 3 where we evaluate the introduced technique using four cases, where we compare the solutions obtained from the SCVS and a standard RBVMS solver. 
The convergence and accuracy of the SCVS are established for a canonical case where an analytical solution is available. 
The computational efficiency of the SCVS is demonstrated and contrasted against that of the RBVMS  in Section 3.6. Finally, we have a look at possible research directions before drawing conclusions at the end in Sections 4 and 5 respectively.

\section{Complex-valued Stokes solver formulation}
In what follows, we first briefly describe the complex-valued Stokes equation as the governing equation for low Reynolds number flows, discuss its discrete formulation using FEM, and then present the SCVS algorithm in detail. 
\subsection{Governing equations}
Consider creeping fluid flow in a domain $\Omega$ with boundary $\Gamma = \partial \Omega$ that is governed by the incompressible unsteady Stokes equations as
\begin{equation}
    \begin{split}
        \rho\frac{\partial \bl u}{\partial t} &= -\nabla p + \nabla \cdot (\mu \nabla \bl u)  \; \; \;  \mathrm {in} \;\Omega, \\
        \nabla \cdot \bl u & = 0 \; \; \;  \mathrm {in} \; \Omega, \\
        \bl u &= \bl g \; \; \; \mathrm {on} \; \Gamma_{\rm g}, \\
        (-p\bl I + \mu\nabla \bl u)\cdot \bl n &= \bl h \; \; \;  \mathrm {on} \; \Gamma_{\rm h},
    \end{split}
    \label{steady_stokes}
\end{equation}
where $\bl u(\bl x,t)$ refers to the velocity of the fluid at location $\bl x$ and time $t$, $p(\bl x,t)$ is the pressure, $\rho$ is the density, $\bl n$ is the boundary $\Gamma$ outward normal vector, and $\mu$ is the viscosity. $\bl g$ and $\bl h$ refer to the imposed velocity and traction on the Dirichlet $\Gamma_{\rm g}$ and Neumann $\Gamma_{\rm h}$ boundaries where $\Gamma = \Gamma_{\rm g} \cup \Gamma_{\rm h}$. 
Additionally, suppose that these boundary conditions are time-periodic with a period of $T$. 

To represent Eq. \eqref{steady_stokes} in the spectral domain, we make use of  
\begin{equation}
    \begin{split}
        \bl u(\bl x,t)=\sum_i {\tilde{\bl u}_i}(\bl x) e^{\hat j \omega_i t},\\
        p(\bl x,t)=\sum_i \tilde{p}_i(\bl x) e^{\hat{j}\omega_i t},
    \end{split}
    \label{vaptransform}
\end{equation} 
where $\hat{j} = \sqrt{-1}$. 
Here the frequency $\omega_i$ is defined in terms of period $T$ as $\omega_i = 2\pi i/T$ for $i=0, 1, \dots , N_{\rm m}$ with $N_{\rm m}$ denoting the largest computed mode. 
For time-periodic flows, such as those encountered in cardiovascular modeling, $T$ is the cardiac cycle, which is the time it takes for boundary conditions to repeat themselves. 
Similar to the state variables, Dirichlet and Neumann boundary conditions can be expressed in the spectral domain as $\bl g(\bl x,t) =\sum_i \tilde{\bl g}_i(\bl x) e^{\hat{j}t\omega_i t}$ and $\bl h(\bl x, t) =\sum_i \tilde{\bl h}_i(\bl x) e^{\hat{j}t\omega_it}$, respectively. 
With these definitions, Eq.~\eqref{steady_stokes} and continuity equation can be written as 
\begin{equation}
    \begin{split}
        \hat{j}\omega_i \rho \tilde{\bl u}_i &= - \nabla \tilde{p}_i + \nabla \cdot (\mu \nabla \tilde{\bl u}_i)  \; \; \; \mathrm {in} \;\Omega, \\
        \nabla \cdot \tilde{\bl u}_i &= 0 \; \; \; \mathrm {in} \; \Omega, \\
        \tilde {\bl u}_i &= \tilde {\bl g}_i \; \; \; \mathrm {on} \; \Gamma_{\rm g},\\
        (-\tilde p_i \bl I + \mu\nabla \tilde {\bl u}_i)\cdot \bl n &= \tilde {\bl h}_i \; \; \; \mathrm {on} \; \Gamma_{\rm h}.
    \end{split}
    \label{stokes_complex}
\end{equation}
These equations resemble those of the steady Stokes equation with a nonzero source term. 
The only difference is that the first (source) term in Eq.~\eqref{stokes_complex} is complex-valued unless $\omega_i = 0$, for which the steady Stokes equations in the real domain are recovered. 
\subsection{Weak formulation}
Since the solution to Eq.~\eqref{stokes_complex} at $\omega_i$ and $\omega_j$ for $i\ne j$ are independent, we only need to provide a solution procedure at a single frequency $\omega_i$.
Thus, for the sake of notation brevity, we drop subscript $i$ and denote our formulation in terms of variable $\omega$ below. 

The weak form of Eq.~\eqref{stokes_complex} is stated as follows. Given the frequency $\omega$, find $\tilde{\bl u} \in \mathcal {\bl S}$ and $\tilde{p} \in \mathcal P$ such that for any $\bl w \in {\bl{\mathcal W}}$ and $q \in \mathcal Q$  
\begin{equation}
    \begin{split}
        B_{\rm G}\left(\bl w, q; \tilde{\bl u}, \tilde{p}\right)  &= F_{\rm G} \left(\bl w, q\right), \\
        B_{\rm G} &=  \int_\Omega \left[ \hat j \omega \rho \bl w\cdot \tilde{\bl u} 
                                             + \nabla \bl w : ( -\tilde{p} I + \mu \nabla \tilde{\bl u}) 
                                             + q \nabla \cdot \tilde{\bl u} \right] \rm d \Omega, \\
        F_{\rm G} &= \int_{\Gamma_{\rm h}} \bl w \cdot \tilde{\bl h} \rm d \Gamma,
    \end{split}
    \label{weak_stokes}
\end{equation}
holds. 
In Eq.~\eqref{weak_stokes}, $\bl w$ and $q$ are test functions for velocity and pressure, respectively, and 
\begin{equation}
    \begin{split}
        \bl{\mathcal S}  &= \left\{ {\tilde{\bl u}} | {\tilde{\bl u}}(\bl x)\in(H^1)^{n_{\rm sd}} ,\; {\tilde{\bl u}} = {\tilde{\bl g}} \;\rm{on}\; \Gamma_{\rm g} \right\}, \\
        \bl{\mathcal W} &= \left\{ \bl w | \bl w(\bl x)\in(H^1)^{n_{\rm sd}} ,\; \bl w = \bl 0 \;\rm{on}\; \Gamma_{\rm g} \right\}, \\
        \mathcal P &= \left\{ \tilde{p} | \tilde{p}(\bl x)\in L^2\right\}, \\
        \mathcal Q &= \left\{ q | q(\bl x)\in L^2 \right\}.
    \end{split}
\end{equation}
 
are velocity and pressure solution and test function spaces. 
$L^2$ denotes the space of scalar-valued functions that are square-integrable on $\Omega$, and  $(H^1)^{n_{sd}}$ denotes the space of vector-valued functions with square-integrable derivatives on $\Omega$.
 
\subsection{Discrete formulation}
Denoting the finite-dimensional subspace of $\bl{\mathcal S}$, $\bl{\mathcal W}$, $\mathcal P$, and $\mathcal Q$ by $\bl{\mathcal S}^h$, $\bl{\mathcal W}^h$, $\mathcal P^h$, and $\mathcal Q^h$, respectively, we attempt to solve the discrete form of Eq.~\eqref{weak_stokes} using Galerkin's approximation. 
Namely, we seek $\tilde {\bl u}^h \in \bl {\mathcal S}^h$ and $\tilde p^h \in \mathcal P^h$ such that for any $\tilde {\bl w}^h \in \bl {\mathcal W}^h$ and $\tilde q^h \in \mathcal Q^h$
\begin{equation}
B_{\rm G}\left(\bl w^h, q^h; \tilde{\bl u}^h, \tilde{p}^h\right)  = F_{\rm G} \left(\bl w^h, q^h\right),
\label{discrete_weak}
\end{equation}
holds. 
In writing Eq.~\eqref{discrete_weak}, we assumed $\Omega^h=\Omega$, i.e., the computational domain after discretization remains unchanged. 
If the two differ, then one must perform the integrals over $\Omega^h$ rather than $\Omega$ when computing $B_{\rm G}$ and $F_{\rm G}$ in Eq.~\eqref{discrete_weak}. 

Galerkin's formulation of incompressible flow has a saddle-point nature, which produces a singular system if one were to adopt similar shape functions for velocity and pressure. 
Various techniques have been proposed to overcome this issue, ranging from mixed-element \cite{brezzi1990discourse} and penalty techniques \cite{hughes1979finite} to stabilized finite element methods \cite{tezduyar1991stabilized}. 
In the present study, we adopt the mixed finite element method, which allows us to reduce the number of variables that influence the accuracy of the SCVS formulation, thus simplifying the measurement of its numerical properties. 
More specifically, we employ linear and quadratic shape functions for  pressure and velocity,  respectively, to satisfy inf-sup condition (also known as LBB condition) \cite{ladyzhenskaya1969mathematical, babuvska1971error, brezzi1974existence}. 
In 2D and 3D, we use linear triangular and tetrahedral elements with 3 and 4 nodal points, respectively, for pressure.
For velocity, we use quadratic shape functions that produce 6 and 10 nodes in each of the aforementioned elements in 2D and 3D, respectively. 
Denoting these linear and quadratic shape functions at global node $A$ by  $N_A(\bl x)$ and $M_A(\bl x)$ , respectively, the test functions and unknowns are interpolated in space using
\begin{equation}
    \begin{split}
        \bl w^h (\bl x) &= \sum_{A\in \eta\setminus\eta_{\rm g}} M_A(\bl x) \bl W_A, \\
        q^h(\bl x) &= \sum_{A\in \hat \eta} N_A(\bl x) Q_A, \\
        {\tilde{\bl u}}^h(\bl x) &= \sum_{A\in \eta \setminus \eta_{\rm g}} M_A(\bl x) \bl U_A + \sum_{A\in \eta_{\rm g}} M_A(\bl x) \bl G_A, \\
        \tilde{p}^h(\bl x) &= \sum_{A\in \hat \eta} N_A(\bl x) P_A,
    \end{split}
    \label{shape_fun}
\end{equation}
where $\eta$, $\eta_{\rm g}$, and $\hat \eta$ refers to the velocity nodes, velocity nodes on the Dirichlet boundaries, and pressure nodes, respectively. 
$\bl U_A$, $P_A$, $\bl W_A$, $Q_A$ in Eq.~\eqref{shape_fun} are the velocity and pressure unknowns and their respective test functions. 
$\bl G_A$ is the prescribed velocity defined on the Dirichlet boundaries after discretization such that
\begin{equation}
        \tilde{\bl g}^h (\bl x) = \Pi^h \tilde{\bl g}(\bl x) =  \sum_{A\in \eta_{\rm g}} M_A(\bl x) \bl G_A,
    \label{dir_dis}
\end{equation}
where $\Pi^h \tilde{\bl g}$ is an operator that projects $\tilde{\bl g}$ to the finite-dimensional discrete space. 

Substituting for the variables appearing in Eq.~\eqref{discrete_weak} using Eq.~\eqref{shape_fun} while ensuring the results holds for any $\bl W_A$ and $Q_A$ produces the following system of linear equations
\begin{equation}
	\bl A \bl X = \bl R,
	\label{linear_sys}
\end{equation}
where
\begin{equation}
	\bl A  = \begin{bmatrix}
	    \bl K & \bl D \\
	    \bl D^{\rm T} & \bl 0
	\end{bmatrix},\;\;\; 
	\bl X = \begin{bmatrix}
	    \bl U \\
	    \bl P
	\end{bmatrix}, \;\;\; 
	\bl R  = 
	\begin{bmatrix}
	    \bl B \\
	    \bl 0
	\end{bmatrix},
	\label{linear_sys_def}
\end{equation}
and $\bl K$ and $\bl D$ matrices and $\bl B$ vector are computed as
\begin{equation}
    \begin{split}
        \bl K_{AB} &= \int_\Omega \left( \hat j \omega \rho M_A M_B + \mu \nabla M_A \cdot \nabla M_B \right) \bl I \rm d \Omega, \\
        \bl D_{AB} &= - \int_\Omega \nabla M_A N_B \rm d \Omega, \\
        \bl B_A &= \int_{\Gamma_{\rm h}} M_A {\tilde{\bl h}} \rm d \Gamma - \bl K_{AB} \bl G_B. 
    \end{split}
    \label{LS_construction}
\end{equation}
The solution to this linear system is obtained using a brute-force approach.
Namely, we treat the entire linear system as a single sparse matrix and solve it iteratively using the Generalized minimal residual method (GMRES) method \cite{saad1986gmres}  with a Jacobi preconditioner. 
All these methods are implemented in our in-house linear solver that employs an optimized data structure for scalable distributed parallel processing \cite{esmaily2015impact}.
The GMRES algorithm used for this purpose is slightly modified from its real counterpart owing to $\bl K$ being complex-valued. 
The same linear solver and preconditioner are also used for the RBVMS to allow for a one-to-one comparison of computational cost between the two. 
This choice has been made due to the underlying structure of the tangent matrix, which is not symmetric for the RBVMS, as well as its relative simplicity and wider use by the finite element community. 
Provided that the left-hand side matrix in the case of SCVS is symmetric and has a block structure, one may adopt a linear solution and preconditioning strategy based on Schur complement to reduce the overall cost of solving this linear system in the future \cite{gresho1998incompressible, olshanskii2007pressure, esmaily2013new, esmaily2015bi}.
The use of multigrid \cite{esmaily2018scalable} is also expected to significantly reduce the cost of linear solve, particularly on finer grids.

\subsection{SCVS algorithm}
In this section, we discuss the implementation of the SCVS algorithm. 
Overall, the procedure involves representing the boundary conditions in the spectral domain, then solving for flow at each mode, and reconstructing the solution in the real domain. 

\begin{enumerate}
\item Take the Fourier transformation of the boundary conditions $\bl g(\bl x,t)$ and $\bl h(\bl x,t)$ as
\begin{equation}
\begin{split}
    \tilde{\bl g}_i(\bl x) &= \frac{1}{T}\int_0^T \bl g(\bl x,t) e^{-\hat{j} \omega_i t} {\rm d}t, \\
    \tilde{\bl h}_i(\bl x) &= \frac{1}{T} \int_0^T\bl h(\bl x,t) e^{-\hat{j} \omega_i t} {\rm d}t. \\
\end{split}
\label{SCVS_bc}
\end{equation}

\item Select the number of modes to truncate the above series based on the smoothness of the boundary conditions.

Note that the number of required modes can be determined directly from the frequency content of the prescribed boundary conditions due to the linearity of the Stokes equations.

More specifically, one can select a tolerance $\epsilon_{\rm m}$ to obtain a priori estimate of the number of required modes $N_{\rm m} + 1$ such that
\begin{equation}
    e^2_{\rm M} = \frac{\|\bl h - \hat{\bl h}\|^2_{L_2(\Gamma_{\rm h}\times T)}}{\|\bl h\|^2_{L_2(\Gamma_{\rm h}\times T)}} + \frac{\|\bl g - \hat{\bl g} \|^2_{L_2(\Gamma_{\rm g}\times T)}}{\|\bl g\|^2_{L_2(\Gamma_{\rm g}\times T)}} < \epsilon^2_{\rm m},
    \label{mode_error}
\end{equation}
in which $\hat{\bl h}(\bl x,t) = \sum_{i=0}^{N_{\rm m}} \tilde {\bl h}_i(\bl x) e^{\hat j \omega_i t}$ and $\hat{\bl g}(\bl x,t) = \sum_{i=0}^{N_{\rm m}} \tilde {\bl g}_i(\bl x) e^{\hat j \omega_i t}$ are the truncated Neumann and Dirichlet boundary conditions using $N_{\rm m}+1$ Fourier modes and $\|\bl h \|_{L_2(\Gamma_{\rm h}\times T)}^2 = \int_{t=0}^T \sum_{\Gamma_i \in \Gamma_{\rm g}}\|\bl h \|_{L_2(\Gamma_i)}^2 {\rm d} t$ and $\|\bl g \|_{L_2(\Gamma_{\rm g}\times T)}^2 = \int_{t=0}^T\sum_{\Gamma_i \in \Gamma_{\rm g}}\|\bl g \|_{L_2(\Gamma_i)}^2 {\rm d} t$.

Even though we are selecting the first $N_m$ modes to truncate the series in this study, in general one may select a non-consecutive set of modes solely based on their amplitude to accelerate the rate of convergence.

\item Construct and solve the linear system in Eq.~\eqref{linear_sys} for $\omega_0$, $\omega_1$, $\dots$, and $\omega_{N_{\rm m}}$ given the boundary conditions computed from the previous step, i.e., $\tilde{\bl g}_i(\bl x)$ and $\tilde{\bl h}_i(\bl x)$.

\item Given the solution $\bl U_A$ and $P_A$, reconstruct the solution in the spectral domain using Eq.~\eqref{shape_fun} and then in time as
\begin{equation}
    \begin{split}
        \bl u^h(\bl x,t)=\sum_{i=0}^{N_{\rm m}} {\tilde{\bl u}^h_i}(\bl x) e^{\hat j \omega_i t},\\
        p^h(\bl x,t)=\sum_{i=0}^{N_{\rm m}} \tilde{p}^h_i(\bl x) e^{\hat{j}\omega_i t},
    \end{split}
    \label{sol_reconstruct}
\end{equation} 
\end{enumerate}

The number of computed modes $N_{\rm m}+1$, which was selected based on a priori error estimate, can be refined following the computation of each solution mode. 
This way, one can compute the difference between the solutions computed from Eq.~\eqref{sol_reconstruct} using $N_{\rm m}-1$ and $N_{\rm m}$ modes and decide whether to continue computing the next mode if that difference is larger than a prespecified tolerance. 

This flexibility presents a significant advantage over the conventional spatio-temporal techniques where capturing an additional mode requires repeating the entire simulation.

Later in Section~\ref{sec:clinical_case}, we will show the rate at which a priori (Eq.~\eqref{mode_error}) and a posteriori (Eq.~\eqref{total_error}) errors drop as $N_{\rm m}$ increases are the same, implying that one can simply rely on Eq.~\eqref{mode_error} to select $N_{\rm m}$. 
In our experience, $N_{\rm m} \le 7$ is typically sufficient for cardiovascular applications where the boundary conditions are smooth and periodic in time. 

\subsection{Error estimation} \label{sec: err_est}
To ensure consistency and accuracy of the proposed formulation, we will consider a set of canonical cases where an analytical solution to the governing equations is available. 
The overall error of a conventional scheme stems from spatial discretization error, the linear solver error, time integration error, and in the case of nonlinear equations, Newton-Raphson iteration error. 
If we consider the solution at a single mode, i.e., boundary conditions oscillating at a single frequency, then SCVS is only prone to the first two types of error enumerated above (discretization and linear solver). 
Thus, in what follows, we establish the overall accuracy of the SCVS algorithm as the mesh size or linear solver tolerance is varied. 
Also, for scenarios in which the boundary conditions are generic functions of time, we will study the convergence of the SCVS solution as a function of $N_{\rm m}$. 

The overall error $e(t)$ is defined based on the relative difference between the computed velocity $\bl u^h(\bl x,t)$ and the reference velocity $\bl u(\bl x,t)$ over the domain $\Omega$ as
\begin{equation}
    e(t) =   \frac{ \|\bl u(\bl x,t)- \bl u^h(\bl x,t)\|_{L_2(\Omega)}}{\|\bl u(\bl x,t)\|_{L_2(\Omega)}},
    \label{total_error}
\end{equation}
where the $L_2(\Omega)$ norm is defined as $\|\bl u\|_{L_2(\Omega)}^2 = \int_{\Omega}\bl u \cdot \bl u\rm d \Omega$. 
As discussed earlier, this total error can be decomposed to the error due to spatial discretization $e_{\rm H}$ and the error due to the linear solver $e_{\rm L}$. 
Denoting the best approximation of the solution in the finite dimensional solution space $\mathcal {\bl S}^h$ by $\Pi^h \bl u$, $L_2$-norm of the interpolation error for the quadratic shape functions employed here can be estimated as \cite{brenner2007mathematical}
\begin{equation}
    e_{\rm H}(t) = \frac{\|\bl u - \Pi^h \bl u\|_{L_2(\Omega)}}{\|\bl u\|_{L_2(\Omega)}} \leqslant C_1 h^3 \frac{\|\bl u\|_{H^3(\Omega)}}{\|\bl u\|_{L_2(\Omega)}},
    \label{int_error}
\end{equation}
where $C_1$ is a constant and $h$ is the diameter of the smallest element-bounding circle in 2D and sphere in 3D.

The error related to the linear solver $e_{\rm L}$ is produced by the approximate nature of the iterative solution procedure employed in solving Eq.~\eqref{linear_sys}. 
This error can be related to the linear solver tolerance $\epsilon_{\rm L}$ as
\begin{equation}
    e_{\rm L}(t) = C_2 \frac{\|\bl R - {\bl A} \widehat {\bl X}\|_2}{\|\bl R\|_2} \le  C_2\epsilon_{\rm L},
    \label{ls_error}
\end{equation}
in which $\widehat{\bl X}$ is the approximate solution obtained from the iterative linear solver. 
In Eq.~\eqref{ls_error}, the constant $C_2$ depends on  the inverse of tangent matrix $\bl A^{-1}$ from Eq.~\eqref{linear_sys_def}  and thus is independent of the boundary conditions.  
How the overall error $e$ changes with $e_{\rm H}$ and $e_{\rm L}$ for a pipe flow will be investigated in Section~\ref{sec:cylinder_case}.

\section{Results} \label{sec:results}
The solution procedure and four case studies selected for evaluating the SCVS algorithm are discussed in this section.
To establish the relative accuracy and cost of the SCVS, we will compare it against a standard RBVMS solver. 
The details pertaining to the RBVMS algorithm are included in~\ref{app:rbvms}. 
The four test cases considered in this study are (1) a 2D channel, (2) a 2D diverging nozzle, (3) a  3D pipe, (4) a patient-specific model of Glenn procedure with an anastomosis \cite{arbia2014numerical}. 
Since the analytical solution is available for the first and third cases, these two cases will be employed to validate our implementation and evaluate its numerical characteristics. 
While the boundary conditions are periodic and contain a single frequency for those two cases, they are generic functions of time for the case two and four, thus allowing us to study the convergence of solution with regard to $N_{\rm m}$ and its relationship to $e_{\rm M}$. 
At the end of this section, the performance of the SCVS and the RBVMS are compared in terms of total CPU time and wall-clock time to demonstrate the performance of the SCVS.

 The SCVS uses mixed quadratic-linear shape functions for velocity and pressure, respectively, whereas the RBVMS uses linear shape functions for both. 
This incompatibility raises a concern that comparing the two is similar to comparing apples to oranges. 
To address this concern, we have implemented a Stokes solver with mixed shape functions (called MSS) that is identical to the SCVS except for the traditional discretization of the acceleration term in the time domain. 
The formulation of the MSS along with its performance metrics relative to the SCVS and the RBVMS are included in~\ref{app:mss}, where we also explain why the SCVS is solely compared against the RBVMS for the remainder of this article.

\subsection{Solution procedure} \label{sec:sol_procedure}
The 2D and 3D models are initially discretized using triangular and tetrahedral elements. 
For this purpose, we use a combination of Tetgen \cite{si2015tetgen} and Simvascular \cite{updegrove2017simvascular} software. 
An in-house script was developed to generate quadratic elements from the linear triangular and tetrahedral elements through a node insertion process. 
For boundary nodes, special care was taken to ensure that inserted nodes are properly projected to be located on the curved boundaries. 
Since the SCVS simulations are performed using quadratic and the RBVMS using linear elements, we reduce the number of elements for the quadratic meshes such that the total number of degrees of freedom is roughly the same as the linear meshes, thus allowing for a one-to-one comparison between the two. 
A zero initial condition is used for all the RBVMS simulations.
 The generalized-$\alpha$ method is used for time integration of the RBVMS, which is an implicit second-order method \cite{jansen2000generalized}. 
As detailed in \ref{app:rbvms}, despite some similarities with the operator splitting techniques \cite{kim1985application}, velocity and pressure are solved together during each linear solve. 
The time step size for the RBVMS is selected to sufficiently resolve the time variation of boundary conditions.
More specifically, we employ 2,000 time steps for cases with time-varying boundary conditions. 
For the steady simulations, the time step size is selected such that the diffusive Courant–Friedrichs–Lewy (CFL) number $\Delta t \nu /h^2 = 1$.
The time integration is continued for steady cases until the relative residual falls below $10^{-6}$. 
A total of five cycles are simulated to achieve cycle-to-cycle convergence for the RBVMS simulations. 
All the results shown below correspond to the last simulated cycle.

For all cases, the Reynolds number is $\rm{Re} < 10^{-3}$ to ensure nonlinear effects are negligible in the case of RBVMS solver. 
The Reynolds number is defined based on the maximum flow rate through boundaries, the area of the boundary with the maximum flow rate, and the kinematic viscosity $\nu=\mu/\rho$.
We only perform a single Newton-Raphson iteration in the RBVMS simulations, given that the governing equations are linear.

The GMRES method is adopted for solving the resulting linear system for both the SCVS and RBVMS using the same tolerance $\epsilon_{\rm L} = 10^{-6}$.
We used our in-house linear solver for this purpose \cite{esmaily2015bi, esmaily2013new, esmaily2015impact}. 
As discussed earlier, the linear system in the case of the SCVS is complex-valued, requiring us to make a necessary adjustment to the real-valued version of our GMRES solver to accommodate for a complex-valued linear system.
Both the RBVMS and the SCVS solvers are also implemented in our in-house finite element solver, which is written in object-oriented Fortran and parallelized using the MPI library. 

\subsection{2D channel flow} \label{sec:channel_flow}
\begin{figure}[H]
    \centering
        \includegraphics[scale=.4]{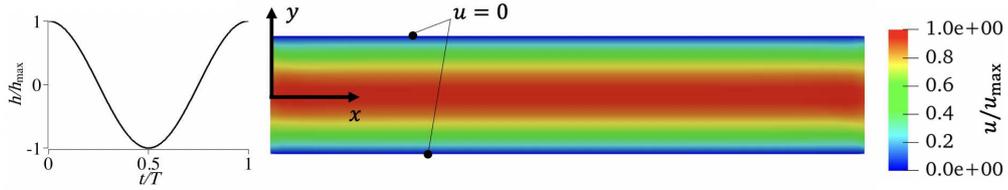}
    \caption{The schematic of the simulated 2D channel flow. 
    A cosinoidal Neumann boundary condition is imposed at the inlet. 
    The shown contour is normalized velocity magnitude at $t=T/4$ for $W=2\pi$ case obtained from the SCVS.}
    \label{fig:channel_geo}
\end{figure}

A 2D channel with a length-to-height aspect ratio of 5 is considered as our first case study (Fig.~\ref{fig:channel_geo}). 
The geometry is discretized using 3,762 linear triangle elements for the RBVMS solver and 882 quadratic elements for the SCVS solver, producing 2,000 and 1,881 nodes and 6,000 and 4,262 degrees of freedom for two cases, respectively. 
The non-slip boundary condition is imposed for the top and bottom walls. 
Either a nonzero constant or a cosinusoidal Neumann boundary condition is imposed on the inlet, and zero traction is imposed at the outlet. These boundary conditions produce a solution that corresponds to the fully developed condition observed in an infinitely long channel (i.e., pressure gradient does not vary as a function of $x$). 
The amplitude of the wave, which otherwise is irrelevant due to the linearity of the Stokes equations, is selected such that the Reynolds number is sufficiently small for the RBVMS computations.  
The oscillation frequency $\omega$ is varied to simulate a wide range of conditions. 
More specifically, the Womersley number $W = \omega H^2/\nu$ with  $H$ denoting the half-channel height varies from 0 to $20\pi$ for the results shown below.

An analytical solution is available for the oscillatory flow in the 2D channel, expressed in the spectral domain as a function of local height $y$ and $\omega$ according to \cite{loudon1998use} 
\begin{equation}
    \tilde u_x(y,\omega) = \left\{ \begin{array}{lr} 
    \displaystyle \frac{\tilde h_x}{2\mu L}(H+y)(H-y), &  \omega = 0, \\
    & \\
    \displaystyle -\frac{\hat{j} \tilde h_x}{\rho L \omega} \left[1-{\cosh}\left(\Lambda\right)^{-1}{\cosh}\left(\Lambda\frac{y}{H}\right)\right], & \omega \neq 0,
    \end{array}\right.
\end{equation}
where $\Lambda^2= \hat{j}W$, $\tilde u_x$ is the streamwise velocity in the spectral domain, $L$ is the channel length, and $\tilde h_x$ is the traction amplitude at $\omega$ imposed at the inlet acting in the streamwise direction. 
This solution can also be expressed in time as
\begin{equation}
    u_x(y,t) = {\rm real}\left\{\tilde u_x(y,\omega) e^{\hat{j} \omega t}\right\}. 
    \label{channel_ana}
\end{equation}

\begin{figure}[H]
    \centering
        \includegraphics[scale=.95]{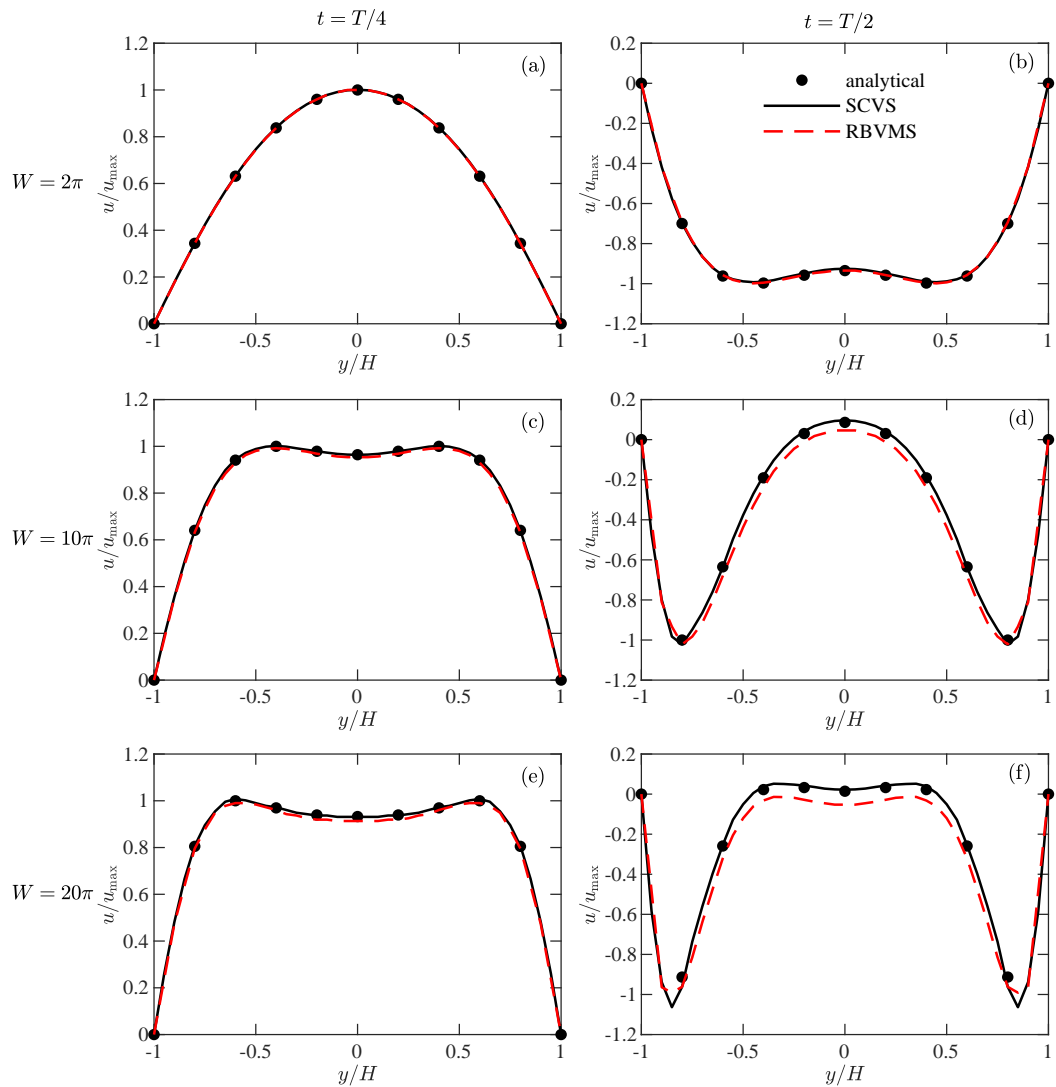}
    \caption{Normalized streamwise velocity as a function of channel height $y/H$ obtained from the SCVS (solid black), the RBVMS (dashed red), and the analytical solution (circles) for the 2D channel flow shown in Fig.~\ref{fig:channel_geo}. 
    The velocity is normalized using analytical $u_{\rm max}$.
    The results on the left and right columns are extracted at $t=T/4$ and $t=T/2$, respectively, and those on the first, second, and third row correspond to $W=2\pi$, $10\pi$, and $20\pi$, respectively.}
    \label{fig:channel_vel}
\end{figure}

The SCVS and the RBVMS simulation results are compared against the analytical solution for $W=2\pi$, $W=10\pi$, and $W=20\pi$ at two time points $t=T/4$ and  $t=T/2$  in Fig.~\ref{fig:channel_vel}. 
In general, the error increases as the flow becomes more oscillatory (at higher $W$). 
This larger error can be attributed to the velocity profile developing sharper gradients near the walls. 
At a similar $W$ and time point, the SCVS provides more accurate predictions in comparison to the RBVMS.
This higher accuracy is despite the fact that a larger number of degrees of freedom were employed in the RBVMS simulations. 
The primary reason for this improved accuracy is the use of quadratic shape function for the SCVS. 
The time integration scheme and the stabilization terms also contribute to the larger errors in the RBVMS results. 

A more condensed version of these results is provided in Table~\ref{tab:channel_err}, where the relative error at time point T/4 and T/2 are computed using Eq.~\eqref{total_error} for the RBVMS and the SCVS. 
In the steady case, since the analytical solution is a parabola and can be exactly represented by the quadratic shape functions, the SCVS error of $e=1.34\times10^{-5}$ is solely due to the linear solver. 
Repeating this computation with a smaller $\epsilon_L$ confirms that $e$ for the SCVS can be arbitrarily lowered without refining the grid.
The steady solution for the RBVMS, on the other hand, shows a relatively larger error of 1.93\% that is grid-dependent given that it utilizes linear shape functions. 
The same improved accuracy is observed at higher modes, where for instance, the RBVMS produces a solution with a 13.4\% error at T/2 and $W=20\pi$, whereas the SCVS error remains as low as 1.83\%. 
In general, in comparison to the RBVMS method, the SCVS is an order of magnitude more accurate for a mesh with a similar number of nodes.

\begin{table}[H]
  \centering
  \begin{threeparttable}
  \caption{Comparison of error in the solution obtained from the SCVS and the RBVMS solver as a function of the Womersley number $W$ computed using Eq.~\eqref{total_error} for the 2D channel flow shown in Fig.~\eqref{fig:channel_geo}. The errors and relative figures are in percent. }
  \label{tab:channel_err}
    \begin{tabular}{ccccccc}
    \toprule
    \multirow{2}{*}{$W = \omega H^2/\nu$}&
    \multicolumn{2}{c}{SCVS (\%)}& \multicolumn{2}{c}{RBVMS (\%)} & \multicolumn{2}{c}{Relative (\%)}\cr
    \cmidrule(lr){2-3} \cmidrule(lr){4-5} \cmidrule(lr){6-7}
    & $e(T/4)$ & $e(T/2)$ & $e(T/4)$ & $e(T/2)$ & $T/4$ & $T/2$ \cr
    \midrule
    
    $0$ & \multicolumn{2}{c}{$1.3\times10^{-3}$} & \multicolumn{2}{c}{$1.9$} & \multicolumn{2}{c}{$6.9\times 10^{-2}$}\cr
    $2\pi$ & $0.01$ & $0.031$ & $0.27$ & $0.40$ & $3.7$ & $7.8$\cr
    $10\pi$ & $0.12$ & $0.46$ & $0.84$ & $8.0$ & $14$ & $5.7$\cr
    $20\pi$ & $0.29$ & $1.8$ & $1.3$ & $13$ & $21$ & $14$\cr
    \bottomrule
    \end{tabular}
    \end{threeparttable}
\end{table}

\subsection{2D diverging nozzle}
\begin{figure}[H]
    \centering
        \includegraphics[scale=.3]{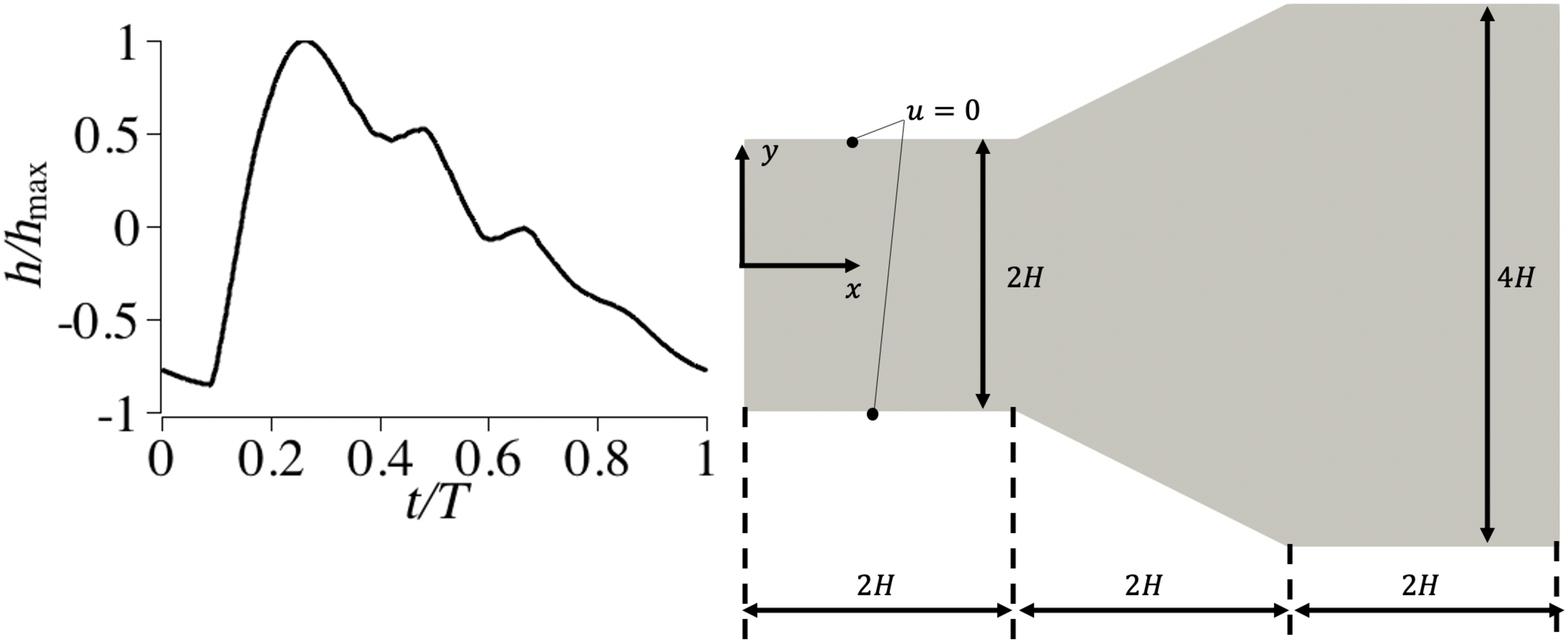}
    \caption{The schematic of the 2D diverging nozzle geometry. 
    A physiologic time-periodic Neumann boundary condition is imposed on the inlet.}
    \label{exp_geo}
\end{figure}

\begin{figure}[H]
    \centering
        \includegraphics[scale=.4]{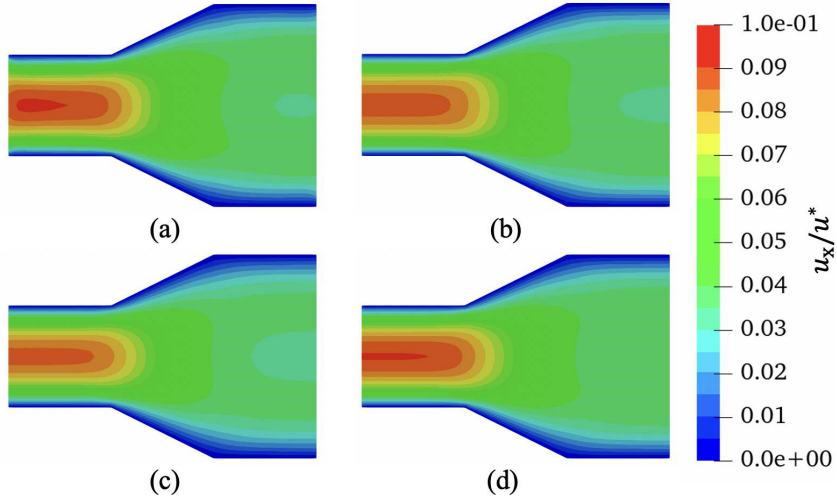}
    \caption{Velocity in the $x$-direction for 2D diverging nozzle shown in Fig.~\ref{exp_geo} at $t=T$ computed using the RBVMS (a) and the SCVS (b-d) with $N_{\rm m} =1$, 3, and 5, respectively.}
    \label{exp_vel}
\end{figure}

In the second example, we consider a 2D diverging nozzle with an expansion ratio of 2 (Fig.~\ref{exp_geo}). 
This geometry is discretized using 2,400 linear triangle elements for the RBVMS solver and 600 quadratic elements for the SCVS solver, producing 1,281 nodes in both cases with 5,124 and 4,184 degrees of freedom, respectively. 
A time-periodic Neumann boundary condition is imposed on the inlet, and zero traction is imposed on the outlet (Fig.~\ref{exp_geo}).
The time variation of the inlet Neumann boundary is selected to resemble a physiologic pressure waveform, with its time average being set to zero to emphasize the role of unsteady modes in the solution. 
The top and bottom walls are considered as non-slip boundaries. 
\begin{figure}[H]
    \centering
        \includegraphics[scale=0.6]{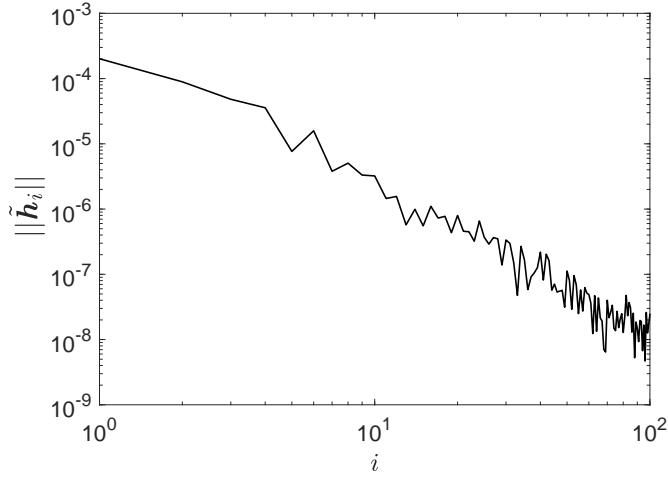}
    \caption{ The frequency content of the Neumann boundary condition imposed at the inlet of 2D diverging nozzle case shown in Fig. \ref{exp_geo}. }
    \label{h_omega}
\end{figure}
 Even though the inlet Neumann boundary contains a wide range of frequencies, only the first few modes have a large enough amplitude to have a significant effect on the results.  
As shown in Fig.~\ref{h_omega}, $\|\tilde {\bl h}_i\|$ follows a power-law trend versus $\omega_i$, experiencing a sharp drop that is approximately proportional $\omega_i^{-2}$.
Such a sharp drop in the amplitude at higher frequencies justifies the truncation of the series at a relatively low wavenumber. 
Thus, to construct the SCVS solution, we select $N_{\rm m}=5$ and solve for $\omega_1, \dots, \omega_5$ while neglecting $\omega_0$ given the time-average of the inlet boundary condition is zero.  
The Womersley number, i.e., $W=\omega H^2/\nu$ with $H$ being the half-inlet height, for these simulated modes ranges from 0 to $10\pi$. 
To show the convergence of SCVS solution with $N_{\rm m}$, we reconstructed solution at $t=T$ using one ($N_{\rm m}=1$), three ($N_{\rm m}=3$), and five ($N_{\rm m}=5$) modes. 
The results of these computations normalized by $u^* = |\tilde {h}_x(\omega=1)| H/\mu$ for $N_{\rm m}=1$, 3, and 5 are shown in Fig.~\ref{exp_vel} along with the results obtained from the RBVMS. 
This figure shows the qualitative convergence of the SCVS at $N_{\rm m}=3$. 

\begin{figure}[H]
    \centering
        \includegraphics[scale=.8]{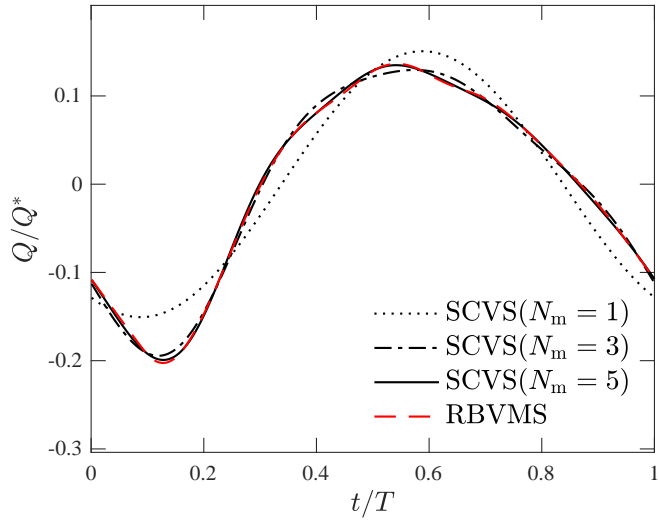}
    \caption{Normalized flow rate for the case shown in Fig.~\ref{exp_geo} obtained from the RBVMS (dashed red) and the SCVS with $N_{\rm m}=1$ (dotted black), 3 (dashed-dot black), and 5 (solid black). }
    \label{exp_flux}
\end{figure}

For a more quantitative comparison of two formulations, the dimensionless flow rate $Q/Q^*$, with $Q^* = Hu^*$ being the characteristic flow rate, is computed and shown in Fig.~\ref{exp_flux} for the RBVMS and the SVC with $N_{\rm m}=1$, 3, and 5. 
Taking the flow computed from the SCVS solution with $N_{\rm m}=10$ as the reference, the relative error in the prediction of the RBVMS is 1.0\%, whereas the relative error for the SCVS at $N_{\rm m}=1$, 3, and 5 is 25\%, 5.0\%, and 1.5\%, respectively.
The fast rate of convergence of the SCVS with respect to $N_{\rm m}$ confirms our earlier argument that only a few modes are needed to obtain a converged solution using the SCVS algorithm when boundary conditions vary smoothly in time. 

\subsection{3D Pipe flow }\label{sec:cylinder_case}
\begin{figure}[H]
    \centering
        \includegraphics[scale=.55]{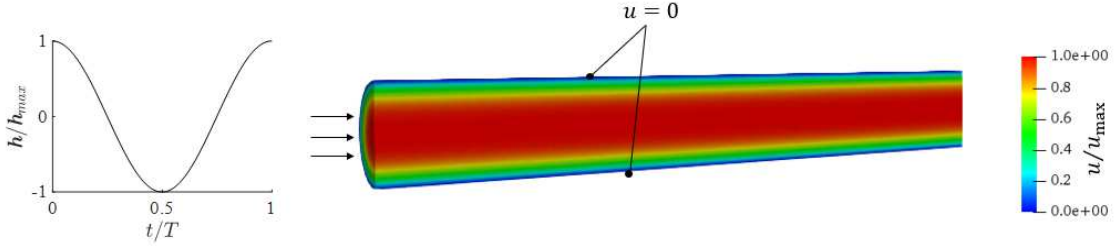}
    \caption{Schematic of the simulated oscillatory laminar flow in a pipe. 
    $h(t)/h_{\rm max}$ is the time trace of the imposed Neumann boundary condition at the inlet. 
    The contour of normalized radial velocity magnitude is shown for $W=8\pi$ case at $t=T/4$ obtained from the SCVS solver.}
    \label{cylinder_geo}
\end{figure}

For the third case study, we consider an oscillatory laminar pipe flow. 
A pipe  with an aspect ratio of $L/R=15$ is considered with a cosinusoidal inlet and zero outlet Neumann boundary condition (Fig.~\ref{cylinder_geo}). 
 Similar to the 2D channel flow case in Section \ref{sec:channel_flow}, these boundary conditions produce a solution that corresponds to the fully developed condition observed in an infinitely long pipe. 
The oscillation period is varied to simulate flow at eleven Womersley numbers $W=\omega R^2/\nu = 0, 8\pi, \dots, 80\pi$ with $R$ denoting the radius of the pipe. 
As we discussed earlier, the domain is discretized in space using linear and mixed quadratic-linear tetrahedral elements for the RBVMS and the SCVS simulations, respectively.
A wide range of meshes is constructed to investigate the convergence of both solvers (Table~\ref{tab:cylinder_msh}). 
The number of elements in corresponding meshes is selected such that the number of nodes is roughly the same for both solvers. 
Overall, the element size normalized by the pipe radius varies from 0.0340 to 0.228 among simulated cases. 

\begin{table}[H]
  \centering
  \begin{threeparttable}
  \caption{Meshes used for discretization of the 3D pipe flow   case shown in Fig.~\ref{cylinder_geo}. 
  QM and LM refer to mixed quadratic-linear and linear tetrahedral meshes, which are employed in the SCVS and the RBVMS simulations, respectively. 
  $N_{\rm ele}$, $N_{\rm nds}$, and $N_{\rm dof}$ denote the numbers of elements, nodes, and degrees of freedom, respectively. }
  \label{tab:cylinder_msh}
    \begin{tabular}{ccccccccccc}
    \toprule
    \multirow{2}{*}{Mesh}&
    \multicolumn{4}{c}{SCVS}&\multicolumn{4}{c}{RBVMS}\cr
    \cmidrule(lr){2-5} \cmidrule(lr){6-9}
    & QM1 & QM2 & QM3 & QM4 & LM1 & LM2 & LM3 & LM4 \cr
    \midrule
    $N_{\rm ele}$ & 24,450 & 49,388 & 95,524 & 162,444 & 207,063 & 374,852 & 728,922 & 1,197,044\cr
    $N_{\rm nds}$ & 37,469 & 70,872 & 133,645 & 225,610 & 37,401 & 64,434 & 122,291 & 197,660\cr
    $N_{\rm dof}$ & 113,193 & 221,894 & 418,400 & 705,311 & 149,604 & 257,736 & 489,164 & 790,640 \cr
    \bottomrule
    \end{tabular}
    \end{threeparttable}
\end{table}

Similar to the first case considered above, an analytical solution is available for this case that permits us to establish the accuracy of each solver. 
For an oscillatory flow in a pipe, the solution in the spectral domain is expressed as \cite{Womersley1955method} 
\begin{equation}
    \tilde{u}_x(r,\omega) = \left\{ \begin{array}{lr} 
    \displaystyle \frac{\tilde{h}_x}{4\mu L}(R^2-r^2), &  \omega = 0, \\ 
    & \\
    \displaystyle -\frac{\hat{j}\tilde{h}_x}{\rho L \omega} \left[1- J_0(\Lambda)^{-1} J_0(\Lambda\frac{r}{R})\right],  & \omega \neq 0,
    \end{array}\right.
    \label{Womersley_cmplx}
\end{equation}
where $\Lambda^2= -\hat{j}W$ and $J_0$ is the zero order Bessel function of the first kind. 
This solution can be expressed in time using
\begin{equation}
    u_x(r,t) = {\rm real}\left\{\tilde u_x(r,\omega) e^{\hat{j} \omega t}\right\}. 
    \label{pipe_ana}
\end{equation}

\begin{figure}[H]
    \centering
        \includegraphics[scale=0.95]{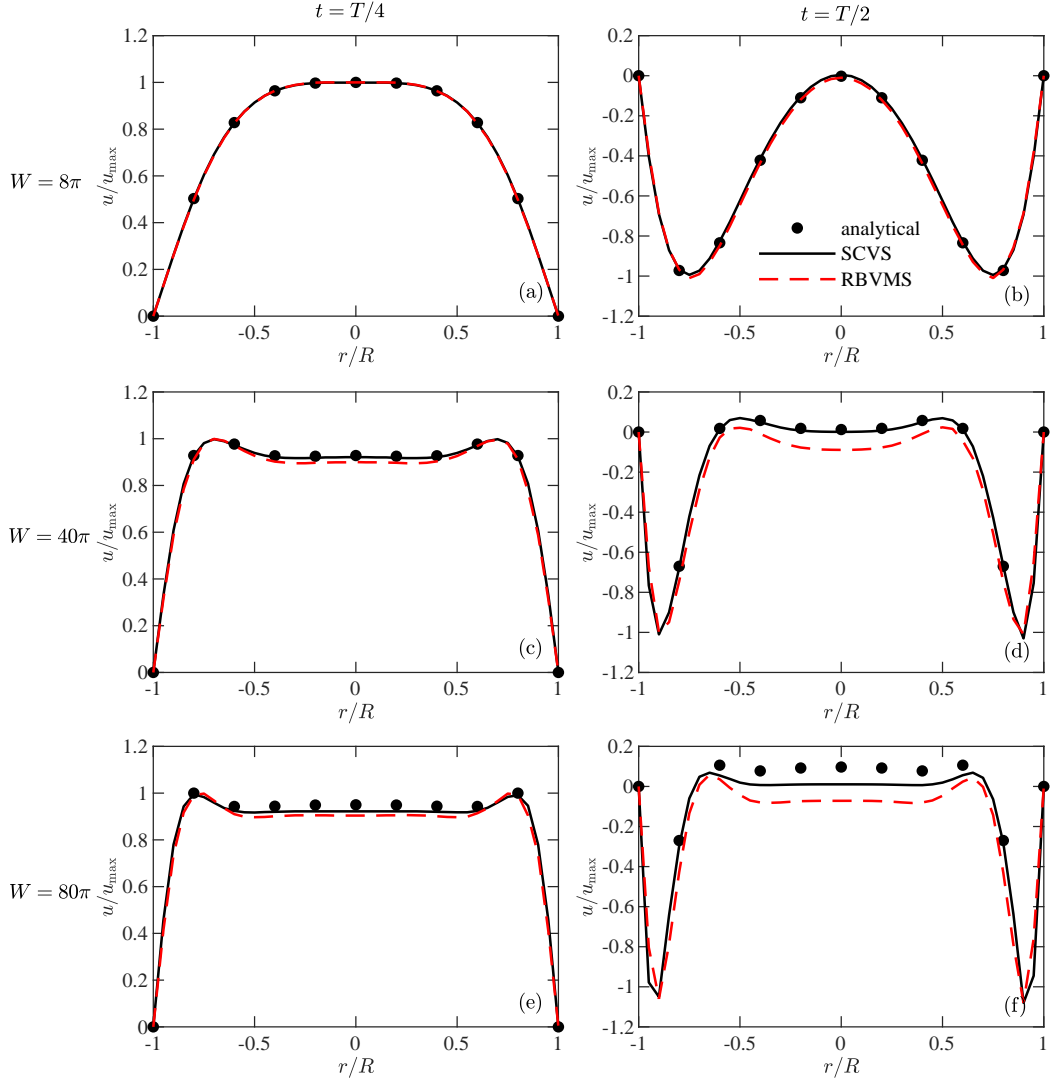}
    \caption{Normalized velocity profiles as a function of radius for an oscillatory laminar pipe flow (shown in Fig.~\ref{cylinder_geo}) predicted using the SCVS (solid black), the RBVMS (dashed red), and analytical solution (circles). 
    The results on the left and right columns are extracted at $t=T/4$ and $t=T/2$, respectively, and those on the first, second, and third row correspond to $W=8\pi$, $40\pi$, and $80\pi$, respectively. 
    Computations are performed on QM4 and LM4 grids for the SCVS and the RBVMS, respectively. 
    All the results are normalized using the maximum velocity from the analytical solution $u_{\rm max}$. }
    \label{cylinder_vel}
\end{figure}

For a qualitative evaluation of the SCVS and the RBVMS solutions, simulated $u_x(r,T/4)$ and $u_x(r,T/2)$ along with the analytical prediction of Eq.~\eqref{pipe_ana} at $W=8\pi$, $40\pi$, and $80\pi$ are shown in Fig.~\ref{cylinder_vel}. 
These results are obtained using the finest grids, namely QM4 and LM4 in Table~\ref{tab:cylinder_msh}. 
Similar to what we observed earlier in Section~\ref{sec:channel_flow} for the 2D channel flow, the SCVS provides a better approximation, particularly at larger Womersely numbers or at $t=T/2$, in which the solution exhibits sharper gradients. 
The error in the steady state solution ($W=0$) obtained from the RBVMS and SCVS solver using the finest grids is $e = 9.0\times10^{-3}$ and $e = 1.4\times 10^{-4}$, respectively. 
The superior accuracy of the SCVS, which is achieved despite using fewer degrees of freedom, is primarily a result of utilizing a higher-order shape function. 
The numerical integration and stabilization terms are the secondary contributors to the larger error in the case of the RBVMS. 

\begin{figure}[H]
    \centering
        \includegraphics[scale=.7]{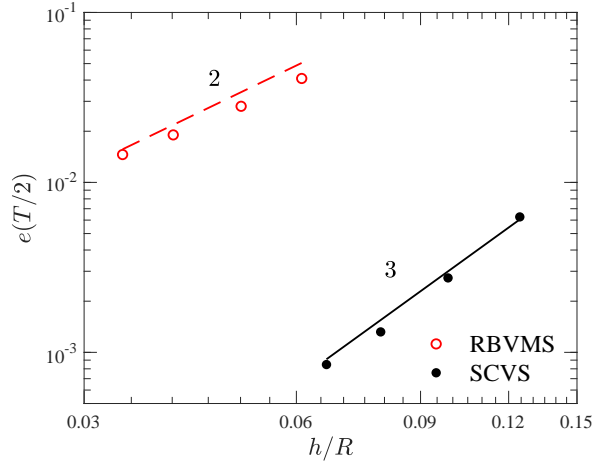}
    \caption{The relative error $e$ (defined in Eq.~\eqref{total_error}) at $t=T/2$ as a function of element size for the oscillatory pipe flow shown in Fig.~\ref{cylinder_geo} at $W=8\pi$. 
    The shown data correspond to the SCVS (solid black circles), the RBVMS (hollow red circles), the analytical estimate of the SCVS error from Eq.~\eqref{int_error} (solid black line), and the analytical estimate of the RBVMS error from Eq.~\eqref{rbvms_error} (dashed red line).}
    \label{cylinder_msh_elesize}
\end{figure}

To study mesh convergence, the error $e(T/2)$ defined in Eq.~\eqref{total_error} is computed for the SCVS and the RBVMS as a function of $h$ at $W=8\pi$ (Fig.~\ref{cylinder_msh_elesize}). 
Third-order accuracy is observed for the SCVS, indicating that the measured error is dominated by the discretization error $e_{\rm H}$. 
This third order accuracy is in agreement with the estimate in Eq.~\eqref{int_error}, which also predicted $e_{\rm H} \propto h^3$. 
A similar relationship can be obtained for the RBVMS, which utilizes linear shape functions, as
 \begin{equation}
     e_{\rm H} \leqslant C_3 h^2 \frac{\|\bl u\|_{H^2(\Omega)}}{\|\bl u\|_{L_2(\Omega)}}. 
     \label{rbvms_error}
 \end{equation}
The second-order accuracy of RBVMS is also confirmed by the results shown in Fig.~\ref{cylinder_msh_elesize}. 

\begin{figure}[H]
    \centering
        \includegraphics[scale=.7]{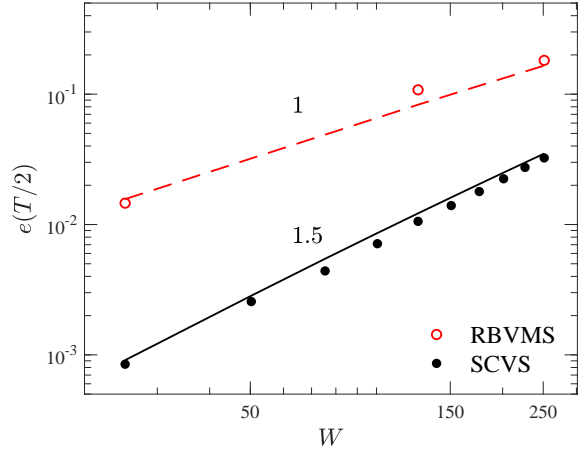}
    \caption{The relative error $e$ (defined in Eq.~\eqref{total_error}) at $t=T/2$ as a function of the Womersley number $W$ for the oscillatory pipe flow shown in Fig.~\ref{cylinder_geo}. 
    The shown results correspond to the SCVS using QM4 grid (solid black circles), the RBVMS using LM4 grid (hollow red circles), the analytical estimate of the SCVS error from Eqs.~\eqref{int_error} and~\eqref{u_norms_ratio} (solid black line), and analytical estimate of the RBVMS error from Eqs.~\eqref{rbvms_error} and~\eqref{u_norms_ratio} (dashed red line).}
    \label{cylinder_msh_freq}
\end{figure}

To further analyze each method's accuracy for various flow conditions, we next study the overall error as a function of the Womersley number $W$. 
As we observed earlier in  Figs.~\ref{fig:channel_vel} and~\ref{cylinder_vel}, the velocity profile develops sharper gradients near the wall at higher modes, thereby producing a larger error at higher $W$. 
That qualitative observation is confirmed quantitatively in Fig.~\ref{cylinder_msh_freq} that shows a direct power-law relationship between $e$ and $W$. 
This relationship indicates that resolving the velocity field at higher modes requires improved spatial resolution. 
It also shows that the quality of the solution degrades faster for quadratic elements than the linear elements as $W$ increases. 
In what follows, we explain these observations in more detail and analytically predict the exponents that appear in the power-law relationship between $e$ and $W$ for the RBVMS and SCVS. 

\begin{figure}[H]
    \centering
        \includegraphics[scale=.7]{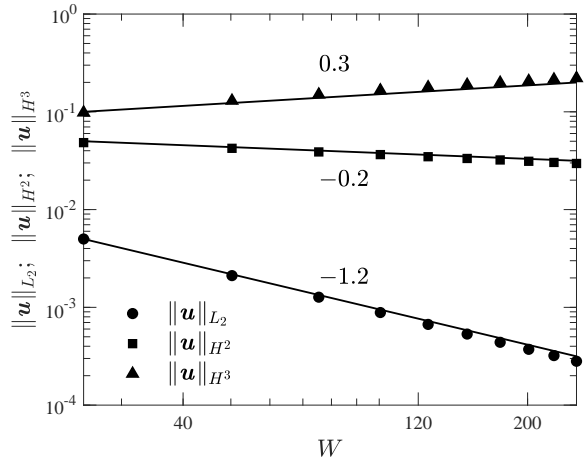}
    \caption{$L_2$ (circles), $H^2$ (squares), and $H^3$ (triangles) norms of the analytical solution $\bl u$ for an oscillatory flow in cylinder as a function Womersley number $W$. 
    The symbols are obtained by 1D numerical integration of Eq.~\eqref{u_norms} whereas the lines with given slopes are curve fits. }
    \label{cylinder_pd}
\end{figure}

Since the linear solver tolerance $\epsilon_{\rm L}$ is selected to be sufficiently small in this case and $e_{\rm L}$ is negligible in comparison to $e_{\rm H}$, we can explain $e\propto W^{3/2}$ for the SCVS and $e\propto W$ for the RBVMS by further analyzing the behavior of $e_{\rm H}$ for each solver.
To utilize Eqs.~\eqref{int_error} and~\eqref{rbvms_error} for this purpose, we first need to establish how $\|\bl u\|_{L_2}$, $\|\bl u\|_{H^2}$, and $\|\bl u\|_{H^3}$ vary with $W$. 
Considering only cases in which $W\ne 0$ and the fact that $u_x = u_x(r,t)$, from Eq.~\eqref{Womersley_cmplx} we can write
\begin{equation}
\begin{split}
    \|\bl u\|_{L_2} & = \left(\int_\Omega u_x^2 {\rm d}\Omega \right)^{1/2}  = \frac{\tilde{h}_x}{\rho \omega} \left(\int_{r=0}^R 2\pi r z_1 \bar z_1 {\rm d}r \right)^{1/2}, \\ 
    \|\bl u\|_{H^2} & = \left\|\frac{\partial^2 u_x}{\partial r^2}\right\|_{L_2} =  \frac{\tilde{h}_x}{\rho w}\left(\frac{\Lambda}{R}\right)^2 \left(\int_{r=0}^R 2\pi r z_2 \bar z_2 {\rm d}r \right)^{1/2}, \\
    \|\bl u\|_{H^3} & = \left\|\frac{\partial^3 u_x}{\partial r^3}\right\|_{L_2} =  \frac{\tilde{h}_x}{\rho w}\left(\frac{\Lambda}{R}\right)^3 \left(\int_{r=0}^R 2\pi r z_3 \bar z_3 {\rm d}r \right)^{1/2},
\end{split}
\label{u_norms}
\end{equation}
where $z_1(r; \Lambda) = 1- J_0(\Lambda)^{-1} J_0(\Lambda\frac{r}{R})$, $z_2(r; \Lambda) = \left[J_2(\Lambda\frac{r}{R})-J_0(\Lambda_i\frac{r}{R})\right]/(2J_0(\Lambda))$, \\$z_3(r; \Lambda) = \left[3J_1(\Lambda\frac{r}{R}) - J_3(\Lambda\frac{r}{R})\right] / (4J_0(\Lambda))$, and $\bar z_1$ denotes complex conjugate of $z_1$. 
Neglecting the dependence of these norms on the integrals of $z_1$, $z_2$, and $z_3$, it is straightforward to show that 
\begin{equation}
\begin{split}
    \|\bl u\|_{L_2} & \propto \displaystyle \frac{\tilde{h}_x R^4}{\mu W}, \\ 
    \|\bl u\|_{H^2} & \propto  \frac{\tilde{h}_xR^2}{\mu}, \\
    \|\bl u\|_{H^3} & \propto \frac{\tilde{h}_x R }{\mu} W^{\frac{1}{2}},
\end{split}
\label{u_norms_simplified}
\end{equation}
indicating $\|\bl u\|_{L_2}$, $\|\bl u\|_{H^2}$, and $\|\bl u\|_{H^3}$ should vary proportional to $W^{-1}$, $W^0$, and $W^{1/2}$. 
Comparing these approximate predictions against reference quantities (obtained from 1D numerical integration of Eq.~\eqref{u_norms}) show that all exponents are over-predicted by 0.2 (Fig.~\ref{cylinder_pd}).
The ratio of these norms, however, is correctly predicted using Eq.~\eqref{u_norms_simplified} as 
\begin{equation}
    \frac{\|\bl u\|_{H^2}}{\|\bl u\|_{L_2}}  \propto R^{-1}W, \;\;\;{\rm and} \;\;\; 
    \frac{\|\bl u\|_{H^3}}{\|\bl u\|_{L_2}}  \propto R^{-3}W^{3/2}, 
\label{u_norms_ratio}
\end{equation}
indicating that the RBVMS and the SCVS error should grow proportional to $W$ and $W^{3/2}$ at sufficiently small $\epsilon_{\rm L}$ owing to Eqs.~\eqref{int_error} and~\eqref{rbvms_error}, respectively.
These analytical predictions are in agreement with our numerical results that were shown earlier in Fig.~\eqref{cylinder_msh_freq}. 

\begin{figure}[H]
    \centering
        \includegraphics[scale=.7]{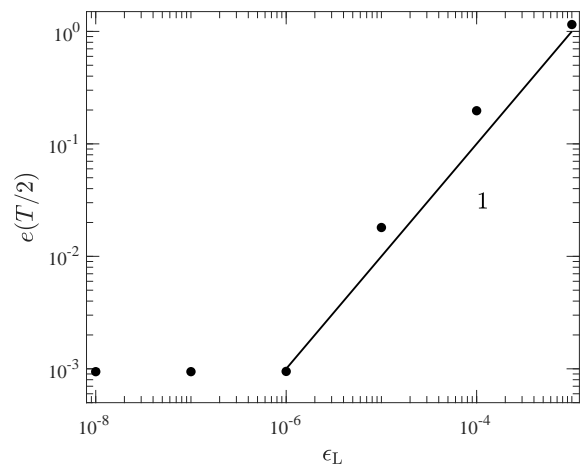}
    \caption{The overall relative error in the SCVS solution $e(T/2)$ as a function of the linear solver tolerance $\epsilon_{\rm L}$.
    The case considered here corresponds to an oscillatory pipe flow shown in Fig.~\ref{cylinder_geo} at $W=8\pi$ and using QM4 mesh.
    A line with a slope 1 is shown for the reference. }
    \label{linear_tols}
\end{figure}

Next, we investigate the effect of linear solver tolerance $\epsilon_{\rm L}$ on the overall accuracy of the SCVS, utilizing QM4 mesh and the solution corresponding to $W=8\pi$ evaluated at $t=T/2$. 
The left and right tail of the results shown in Fig.~\ref{linear_tols} indicate that the overall error $e$ is independent of $e_{\rm L}$ for sufficiently small $e_{\rm L}$ and proportional to $\epsilon_{\rm L}$ when $e_{\rm L}$ is sufficiently large. 
This observation confirms that (as expected) the overall error is dominated by the discretization error $e_{\rm H}$ and that of the linear solver $e_{\rm L}$ at small and large $\epsilon_{\rm L}$, respectively. 
The transition between dominance of $e_{\rm H}$ and $e_{\rm L}$ occurs at approximately $\epsilon_{\rm L} = 10^{-6}$, which is a value specific to this case study and a function $h$ among others.
Note that this tolerance is the optimal tolerance with regard to the computational cost, as further decrease in $\epsilon_{\rm L}$ is not met with an improved overall solution accuracy.
Lastly, note that the slope of 1 observed on the right tail of Fig.~\ref{linear_tols} is in agreement with Eq.~\eqref{ls_error}, which predicts $e_{\rm L} \propto \epsilon_{\rm L}$.  

\begin{figure}[H]
    \centering
        \includegraphics[scale=.7]{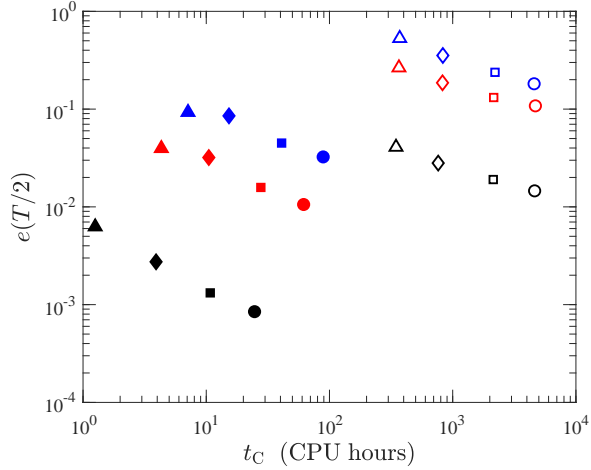}
    \caption{Relative error $e(T/2)$ as a function of the computational cost $t_{\rm C}$ for the 3D pipe flow example shown in Fig.~\ref{cylinder_geo}.
    Symbols with the same color should be compared. 
    The solid symbols correspond to the SCVS and hollow to the RBVMS. 
    The circle, square, diamond, and triangle symbols correspond to QM4/LM4, QM3/LM3, QM2/LM2, and QM1/LM1 meshes listed in Table~\ref{tab:cylinder_msh}. 
    The black, red, and blue colors correspond to $W=8\pi$, $40\pi$, and $80\pi$, respectively. 
    These results show that the SCVS, compared to the RBVMS, always provides a higher accuracy at a lower cost.}
    \label{cpu_time}
\end{figure}

If we take the main objective of designing a computational method to be making a viable prediction at a specified accuracy with the lowest cost possible, then it is desirable to compare the SCVS and the RBVMS methods on an error-versus-cost plot. 
Thus, we have extracted the cost (in terms of CPU time) and error (in terms of $e(T/2)$) for all the simulations performed in this section using various grids and $W$. 
The results are shown in Fig.~\ref{cpu_time}.
Comparing the symbols with the same color, i.e., for a flow at a given $W$, the SCVS always provides more accurate predictions at a lower cost. 
Focusing on $W=80\pi$ cases (blue symbols), the SCVS yields a similar accuracy at the coarsest grid (solid triangle) to that of the RBVMS at the finest grid (hollow circle) while reducing the cost by three orders of magnitude. 
If one holds the computational cost relatively similar (solid blue circle versus hollow blue triangle, for instance), the SCVS offers over an order of magnitude improvement in accuracy relative to the RBVMS. 

\subsection{3D patient-specific Glenn}\label{sec:clinical_case}
\begin{figure}[H]
    \centering
        \includegraphics[scale=.4]{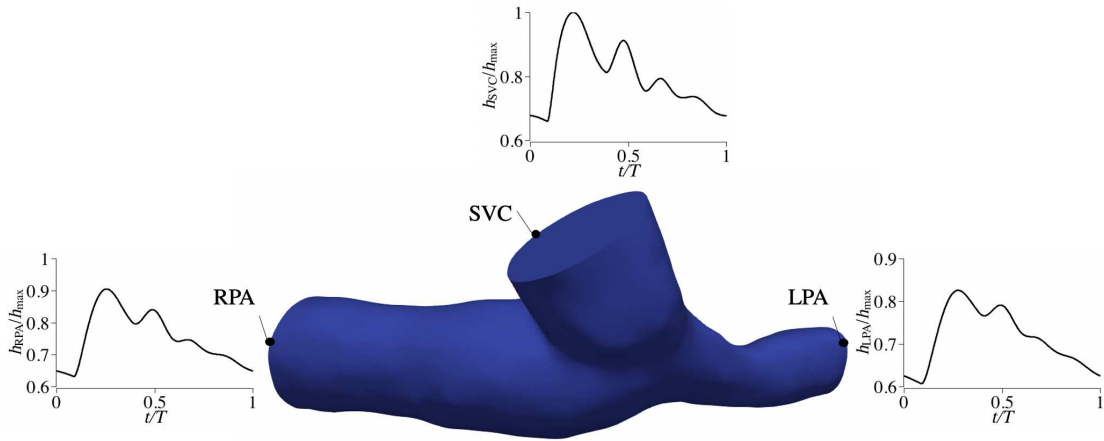}
    \caption{The geometry and boundary conditions employed for the patient-specific case study. $h_{\rm max}$ denotes the maximum value of traction imposed on the SVC.}
    \label{stubby_geo}
\end{figure}

For the last test case, we will consider a complex patient-specific geometry acquired from a patient undergoing Glenn operation \cite{arbia2014numerical}.
As shown in Fig.~\ref{stubby_geo}, the superior vena cava (SVC) is anastomosed to the left pulmonary artery (LPA) and right pulmonary artery (RPA) in this operation.
The geometry is discretized using 988,747 linear tetrahedral elements for the RBVMS solver and 132,066 quadratic tetrahedral elements for the SCVS. 
This discretization results in 163,791 and 183,708 nodes and 655,164 and 574,401 degrees of freedom for the linear and quadratic meshes, respectively. 
The boundary conditions are selected based on physiologic data and a Windkessel model \cite{esmaily2012optimization, vignon2010outflow} for the RPA and LPA faces (Fig.~\ref{stubby_geo}).
The wall is considered a non-slip boundary. 

In this example, twelve modes ($N_{\rm m}=11$) are simulated in total for the SCVS. 
The case with $N_{\rm m} = 11$ is used as the reference to measure error for the remaining cases, given that the solution accuracy of the SCVS was superior to that of the RBVMS based on the cases discussed earlier. 
The Womersley number $W = \omega D_{\rm h}^2/\nu$ is based on the hydraulic diameter of the SVC ($D_{\rm h}$), which is defined such that $\pi D_{\rm h}^2/4$ is equal to the area of the SVC boundary. 
This definition, which is the square of what is normally defined in the literature as Womersley number, leads to $W= 0$, $146\pi$, $293\pi$, $\dots$, $1611\pi$ for simulated $\omega_0$, $\omega_1$, $\omega_2$, $\dots$, $\omega_{11}$.
The results of these computations are normalized by $u^* = |\tilde {h}_{\rm SVC}(\omega=0)| D_{\rm h}/\mu$ and $Q^*=\pi D_h^2 u^*/4$. 

\begin{figure}[H]
    \centering
        \includegraphics[scale=.44]{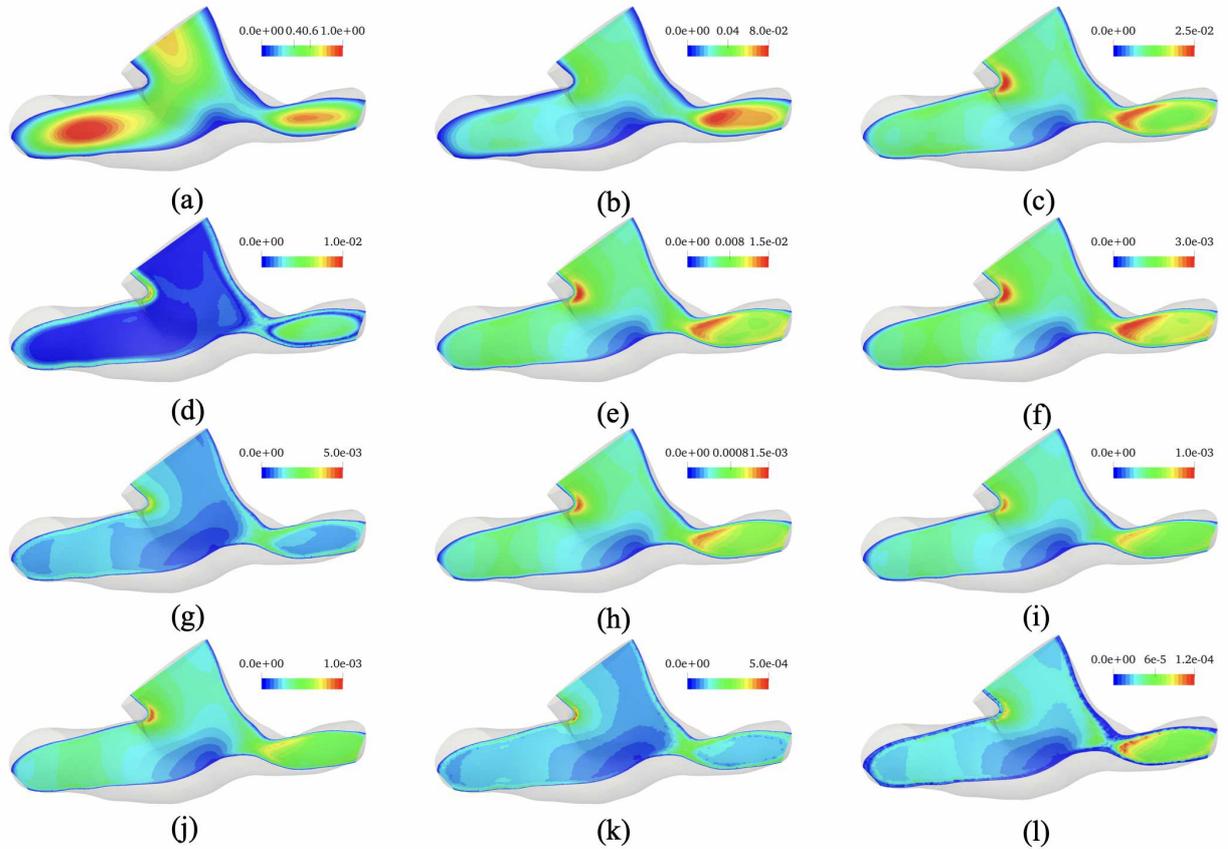}
    \caption{The velocity magnitude normalized by $u^*$ obtained from the SCVS solver at $t=T/2$ for (a) $W = 0$, (b) $W = 146\pi$, (c) $W = 293\pi$, (d) $W = 439\pi$, (e) $W = 586\pi$, (f) $W = 732\pi$, (g) $W=879\pi$, (h) $W=1025\pi$, (i) $W=1171\pi$, (j) $W=1318\pi$, (k) $W=1464\pi$, (l) $W=1611\pi$.}
    \label{stubby_modes}
\end{figure}

A closer examination of the solutions associated with each mode shows that their spatial distribution widely varies as $W$ increases (Fig.~\ref{stubby_modes}). 
While the velocity peaks at the center of the vessels for the steady solution (Fig.~\ref{stubby_modes}-(a)), it develops two peaks closer to the walls at a larger $W$.
This behavior resembles the trend that we observed earlier in the canonical geometries (e.g., cylinder in Section~\ref{sec:clinical_case}). 
This behavior is a result of a change in the relative importance of various terms that appear in the Stokes equation. 
While only the viscous and pressure terms are active at $W=0$, three terms (acceleration, pressure, and viscous) balance each other for $W>0$. 
As $W$ increases, the relative importance of the acceleration term increases, leading to a solution that exhibits peaks near the walls rather than at the center of the geometry. 

\begin{figure}[H]
    \centering
        \includegraphics[scale=.3]{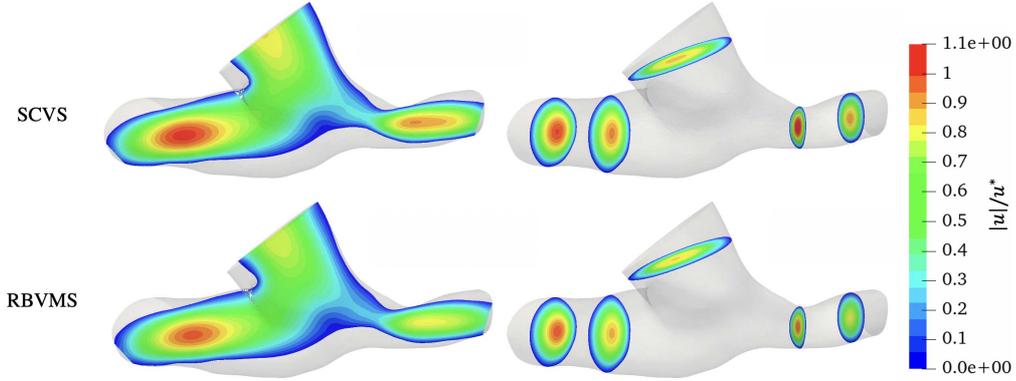}
    \caption{Normalized velocity magnitude at $t=1/2T$ obtained from the SCVS with $N_{\rm m} = 5$ (top) and the RBVMS (bottom) for the case shown in Fig.~\ref{stubby_geo}.}
    \label{stubby_vel}
\end{figure}

The SCVS solution in the time domain is reconstructed using Eq.~\eqref{sol_reconstruct} and qualitatively compared against that of the RBVMS (Fig.~\ref{stubby_vel}). 
This qualitative comparison shows that a few modes (in this case 6, including the steady solution) are sufficient for resolving the solution in this complex geometry. 

\begin{figure}[H]
    \centering
        \includegraphics[scale=.5]{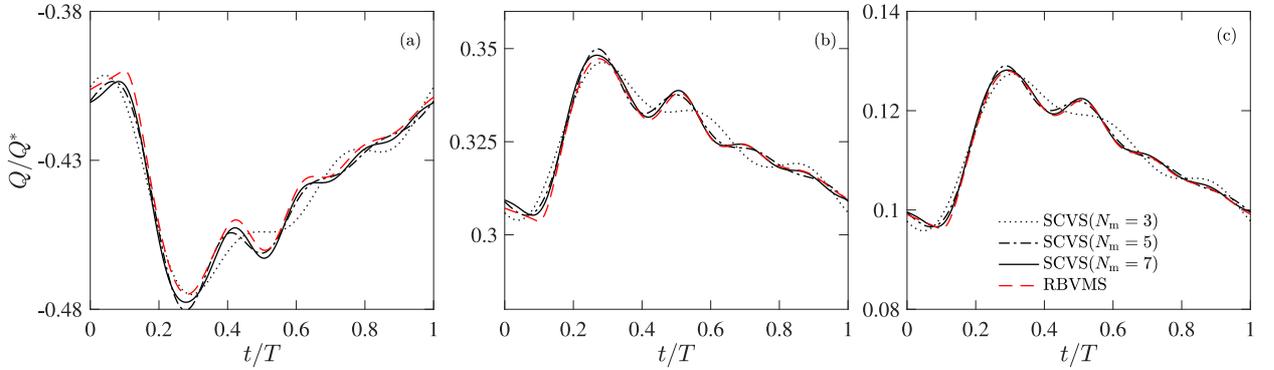}
    \caption{Normalized flow rate $Q/Q^*$ through the SVC (a), LPA (b), and RPA (c) predicted by the RBVMS (dashed red) and the SVC with $N_{\rm m}=3$ (dotted black), $N_{\rm m}=5$ (dashed-dot black), $N_{\rm m}=7$ (solid black) for the case shown in Fig.~\ref{stubby_geo}.}
    \label{stubby_flux}
\end{figure}

For a more quantitative comparison between the SCVS and the RBMVS solutions, the predicted flow through three branches is shown in Fig.~\ref{stubby_flux}. 
This figure confirms that the difference between the two solvers' solution reduces as more modes are employed. 
Using the case with $N_{\rm m}=11$ as a reference, the relative errors of flux computed from the RBVMS are $0.65\%$, $0.38\%$, and $0.34\%$ on the SVC, LPA, and RPA, respectively. 
The relative errors in the SCVS solution with $N_{\rm m}=7$ are $0.14\%$, $0.19\%$, and $0.12\%$ for the SVC, LPA, and RPA, respectively. 
Note that, due to the error in the RBVMS solution (as well as discretization error in the SCVS), the two solutions do not converge even at very large $N_{\rm m}$. 
Nevertheless, the SCVS does converge to a solution as $N_{\rm m} \to \infty$. 
The rate of this convergence depends on the smoothness of the boundary conditions. 

\begin{figure}[H]
    \centering
        \includegraphics[scale=.7]{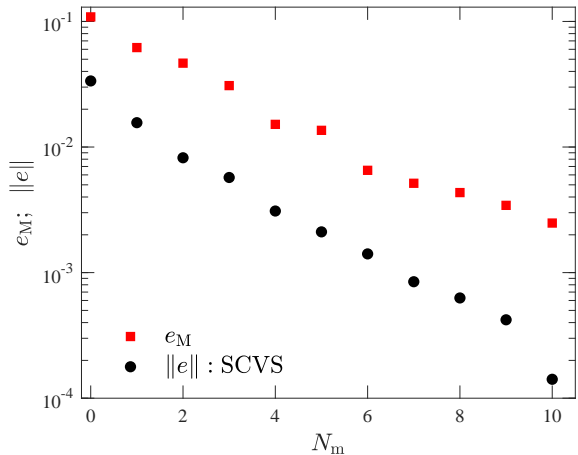}
    \caption{The overall error $\|e(t)\|_{L_2([0,T])}$ (solid circles) as a function of $N_{\rm m}$ for the case shown in Fig.~\ref{stubby_geo}, using the case with $N_{\rm m}=11$ as the reference. 
    The truncation error for the imposed Neumann boundaries $e_{\rm M}$ (red squares) is also shown, which follows the same trend as $\|e\|$. }
    \label{stubby_flux_err}
\end{figure}

The relationship between the smoothness of the boundary condition time variation and the convergence rate of the SCVS with respect to $N_{\rm m}$ is demonstrated in Fig.~\ref{stubby_flux_err}. 
$e_{\rm M}$ from Eq.~\eqref{mode_error}, which measures the truncation error in the imposed boundary conditions associated with the finite $N_{\rm m}$, follows the same trend as the overall error $e$. 
As more modes are included in the solution, the boundary conditions approach their temporal reference profile, thereby improving the overall accuracy of the SCVS solution. 
Note that the two trends slightly diverge on the right tail of this plot. 
This divergence is an artifact of taking the case with $N_{\rm m} = 11$ as the reference when measuring $\|e\|$, whereas $e_{\rm M}$ is measured against the reference solution. 
If the exact solution were available and employed as the reference solution, then $\|e\|$ for very large $N_{\rm m}$ would have converged to the larger of discretization $e_{\rm H}$ and linear solver $e_{\rm L}$ errors rather than going to zero. 

\subsection{Computational costs}
The primary reason for the development of the SCVS algorithm was to obtain a more computationally efficient solution procedure. 
We discussed some of the performance metrics of the SCVS algorithm in the discussion pertaining to Fig.~\ref{cpu_time}. 
In this section, we provide a more comprehensive account of the performance of these two solvers in terms of computational costs as well as wall-clock-time for all cases studied above. 

For cases with multiple grids, we considered the finest grid in computing performance values. 
Also, we consider $N_{\rm m} =5$ for the SCVS as it proved sufficient for sufficiently smooth boundary conditions. 
Thus, the computational cost $t_{\rm C}$ for the SCVS is computed as the cumulative cost of simulating the first 6 modes. 
Its wall-clock-time $t_{\rm W}$, on the other hand, is taken as the maximum of the wall-clock-time of those 6 simulated modes since they are embarrassingly parallelizable. 
The simulation results presented in Sections~\ref{sec:channel_flow} and~\ref{sec:cylinder_case} for the canonical 2D channel and 3D pipe flow   cases were for individual modes. 
Nevertheless, we consider the sum of the cost of the first six modes for those cases as well.
This choice was made to represent a situation in which the flow in these models is simulated with an arbitrary boundary condition rather than one with a unimodal oscillation. 
Furthermore, note that both $t_{\rm C}$ and $t_{\rm W}$ for the RBVMS scale linearly with the total number of time steps\footnote{This statement applies only to the linear equations considered here, since a large $\Delta t$ increases the number of Newton-Raphson iterations at higher Reynolds numbers, thus changing the cost of advancing the solution for a time step.}.
The results reported here are what is considered typical, namely simulating five cycles with each including 2,000 time steps for a total of 10,000 time steps. 
Finally, the performance metrics are in general implementation- and machine-dependent. 
However, given that both the RBVMS and the SCVS solvers are implemented by our group, written and compiled using the same language and compiler, linked against the same libraries, and ran on the same machine, the figures in Tables~\ref{tab:cost} provide an apple-to-apple comparison of the two formulations. 

\begin{table}[H]
  \centering
  \begin{threeparttable}
  \caption{Comparison of the computational performance of the SCVS and RBVMS solvers. 
  $N_{\rm p}$, $t_{\rm C}$, $t_{\rm W}$ denote the number of processors, computational costs in CPU-hours, and the simulation wall-clock-time in hours, respectively. 
  The last two columns are $t_{\rm C}$ and $t_{\rm W}$ for the SCVS relative to those of the RBVMS. 
  All the SCVS results correspond to $N_{\rm m}=5$. 
  For cases with multiple grids, we adopted the finest grid in computing these performance figures.
  The relative figures are in percent. }
  \label{tab:cost}
    \begin{tabular}{lccccccccc}
    \toprule
    \multirow{2}{*}{Case}&
    \multicolumn{3}{c}{SCVS}&\multicolumn{3}{c}{RBVMS}& \multicolumn{2}{c}{Relative (\%)} \cr
    \cmidrule(lr){2-4} \cmidrule(lr){5-7} \cmidrule(lr){8-9}
    & $N_{\rm p}$ & $t_{\rm C}$ (hr) & $t_{\rm W}$ (hr) & $N_{\rm p}$ & $t_{\rm C}$ (hr) & $t_{\rm W}$ (hr) & $t_{\rm C}$ & $t_{\rm W}$ \cr
    \midrule
    2D channel & 6 & $1.5\times10^{-2}$ & $3.9\times10^{-3}$ & 16 & 1.9 & 0.12 & 0.79 & 3.4 \cr
    2D nozzle & 6 & $7.4\times10^{-3}$ & $1.8\times10^{-3}$ & 16 & 1.0 & 0.064 & 0.74 & 2.9\cr
     3D Pipe    & 384 & $2.3\times10^2$ & 1.0 & 64 & $4.6\times10^3$ & 72 & 5.1 & 1.3\cr
    3D Glenn & 768 & $3.2\times 10^2$ & 0.52 & 128 & $3.0\times10^3$ & 23 & 11 & 2.2 \cr
    \bottomrule
    \end{tabular}
    \end{threeparttable}
\end{table}

As was hypothesized earlier, the SCVS reduces the overall computational cost significantly when it is compared against the standard RBVMS formulation. 
This reduction in cost that ranges from less than 1\% for 2D cases to 11\% for the 3D Glenn geometry is primarily achieved by reducing the number of linear solves. 
While this number is $N_{\rm m} + 1 = 6$ for the SCVS, it is equal to the total number of time steps for the RBVMS, i.e., 10,000.
The performance gap between the SCVS and RBVMS is not as large as $1667=10,000/6$ as a single SCVS linear solver is much more expensive than that of the RBVMS. 
On average, the SCVS requires 30,000 GMRES iterations, whereas 500 iterations suffice for solving the linear system obtained for the RBMVS formulation. 
The reason for the large number of iterations for the SCVS is mostly attributed to the high condition number of $\bl A$ matrix that has a zero diagonal sub-block (Eq.~\eqref{linear_sys_def}).  
As discussed in Section~\ref{sec:future}, reducing the cost of solving this linear solver presents an opportunity for significantly reducing the cost of the SCVS algorithm in the future. 

The performance gap between the SCVS and RBVMS widens when wall-clock time is concerned as $t_{\rm W}$ for the SCVS is less than 4\% of that of the RBVMS for all cases considered here (Table~\ref{tab:cost}). 
As we discussed earlier, the modes in the SCVS algorithm are independent and can be solved concurrently. 
Therefore, the number of processors utilized in the SCVS computation can far exceed that of the RBVMS without loss of parallel efficiency.
Differently put, the wall-clock time for the SCVS is roughly independent of the number of computed modes as long as there are sufficient parallel computational resources available. 
The cumulative effect of this added dimension for parallelization and lower overall computational cost is a formulation that allows for a much faster turn-around time for a given problem.


\section{Future work} \label{sec:future}
The biggest hurdle to be overcome in the future is the extension of the SCVS formulation to high Reynolds number flows. 
Such an extension will not be trivial, as one can not exploit the linearity of the governing equations to solve for velocity and pressure at different modes independently. 

Solving for all the modes at the same time will be one way to deal with this coupling. 
However, this brute force approach can become quickly unaffordable as the number of modes increases. 
This limitation, which has been observed in the past for the Time Spectral Method discussed in Section \ref{sec:introduction}, must be overcome if the extension of SCVS to the Navier-Stokes equation were to find widespread use in the future.

The efficiency of the SCVS algorithm can be significantly improved in the future by making minor adjustments in its formulation or its underlying linear solver. 
Relaxing the incompressibility constraint by using a penalty method or including stabilization terms in its formulation is expected to reduce the condition of the underlying linear system significantly. 
Experimenting with more efficient linear solver algorithms (e.g., bi-partitioned method \cite{esmaily2015bi}) as well as preconditioners \cite{esmaily2013new}  can lead to additional improvements in the overall performance of the SCVS algorithm in the future. 

Depending on the hardware architecture, reformulating the problem so that it avoids complex arithmetic may or may not reduce the cost of this scheme as such reformulation leaves the number of real floating-point operations unchanged. 

\section{Conclusions}
In this paper, we proposed the SCVS as an alternative approach for fast simulation of time-periodic flow at low Reynolds numbers in complex geometries. 
Starting from the unsteady Stokes equation, we showed that it could be expressed as a steady Stokes equation with an imaginary source term in the time-spectral domain. 
The resulting equation can then be discretized and solved as a boundary value problem at a few selected modes, avoiding the use of costly and unscalable time integration schemes. 
As a proof of concept, we showed how this boundary value problem could be solved using Galerkin's formulation with mixed elements.
We later employed this formulation for simulating flow in a variety of 2D and 3D geometries. 
To provide a point of reference, all these simulations were also performed using the standard RBVMS formulation.

For cases with an analytical solution available, the SCVS showed about an order of magnitude improvement in accuracy relative to the RBVMS at a similar number of degrees of freedom.
This difference was attributed to the use of quadratic shape functions for the SCVS and to a lesser extent to the lack of stabilization terms and time integrator in our formulation. 
The variation of the overall error in the SCVS solution as a function of grid size, linear solver tolerance, and the mode number was measured through our numerical test cases and was shown to follow analytically-derived power-law formulas. 
For cases with boundary conditions changing arbitrarily in time, we showed the overall error follows the same trend as the truncation error associated with simulating a finite number of modes. 
For smooth time-varying boundary conditions, such as those encountered in cardiovascular flows, we showed that using as few as 6 modes is sufficient for reducing the overall error to $O(10^{-3})$. 

The SCVS led to significant improvement in performance in comparison to the RBVMS. 
The total computational cost of the SCVS for the cases considered in this study varied between 0.74\% and 11\% of that of the RBVMS, whereas its wall-clock time was consistently lower than 4\% of that of the RBVMS. 
Further improvement in performance through the use of stabilized or penalty formulation as well as the extension to higher Reynolds numbers remain to be explored in the future. 

\appendix
\section{RBVMS formulation} \label{app:rbvms}
A brief account of the RBVMS formulation employed in this study is provided below. 
A more detailed description of this algorithm is provided in reference \cite{brooks1982streamline, bazilevs2007variational, esmaily2015bi}. 

The weak formulation of the RBVMS is stated as follows. Find $\bl u \in \bl {\mathcal S}$ and $p \in \mathcal P$, such that for all $\bl w \in \bl {\mathcal W}$ and $q \in \mathcal Q$
\begin{equation}
   B_G\left(\bl w, q; \bl u, p\right) + B_S\left(\bl w, q; \bl u, p\right) = F \left(\bl w, q\right), 
\end{equation}
where
\begin{equation}
    \begin{split}
    B_G & =  \int_\Omega \left[ \rho \bl w \cdot (\dot {\bl u} + \bl u \cdot \nabla \bl u) + \nabla \bl w : ( -p \bl I + \mu \nabla^{\rm s} \bl u ) + q \nabla \cdot \bl u \right] \rm d \Omega \\
    B_S & = \sum_{e\in \bl I_{\rm e}} \int_{\Omega^e} \left[ \rho \nabla \bl w : \left( \bar \tau \bl u_{\rm p} \otimes (\bl u_{\rm p} \cdot \nabla \bl u )
   - \bl u \otimes \bl u_{\rm p} + \tau_{\rm C} \nabla \cdot \bl u \bl I \right) + \rho \bl w \cdot \left( \bl u_{\rm p} \cdot \nabla \bl u \right)
   - \bl u_{\rm p} \cdot \nabla q \right] \rm d \Omega, \\ 
   F & = \int_{\Gamma_{\rm h}} \bl w \cdot \bl h \rm d \Gamma. 
   \end{split}
   \label{rbvms_eq}
\end{equation}
In this equation, $B_G$ contains  Galerkin's term whereas $B_S$ are the stabilization added to allow for equal-order velocity and pressure functions and prevent convective instability associated with Galerkin's method. 
Other parameters appearing in~\eqref{rbvms_eq} are defined as
\begin{equation}
\begin{split}
   \bl u_{\rm p} & = - \tau_{\rm M} \left(\dot {\bl u} + \bl u \cdot \nabla \bl u + \frac{1}{\rho} \nabla p - \frac{\mu}{\rho} \nabla^2 \bl u - \bl f \right), \\
   \tau_{\rm M} & = \left[ \left( \frac{2c_1}{\Delta t} \right)^2 + \bl u \cdot \bl{\xi} \bl u + c_2 \left(\frac{\mu}{\rho}\right)^2 \bl {\xi} : \bl {\xi}\right]^{-\frac{1}{2}}, \\
   \bar \tau & = \left( \bl u_{\rm p} \cdot \bl{\xi} \bl u_{\rm p} \right)^{-\frac{1}{2}}, \\
   \tau_{\rm C} & = \left[ \rm {tr} \left( \bl{\xi} \right) \tau_{\rm M} \right]^{-1}, \\ 
   \end{split}
   \label{stab_param}
\end{equation}
in which $c_1=1$ and $c_2=3$ are the model constants, $\bl {\xi} \in \mathbb R^{n_{\rm sd}} \times \mathbb R^{n_{\rm sd}}$ is covariant tensor obtained from a mapping between the physical and parent domains, and $\Delta t$ is the time step size. 
The resulting equations are discretized using triangular and tetrahedral elements in 2D and 3D.  
To relate $\dot {\bl u}$ to $\bl u$ and integrate the resulting equations in time, the second order generalized-$\alpha$ method is adopted \cite{jansen2000generalized}.
This algorithm, which is implicit in time, requires linearization of the nonlinear terms in Eq.~\eqref{rbvms_eq}.
Thus, we employ the Newton-Raphson method for linearization of the resulting equations and a predictor-corrector procedure to converge on the solution at the next time step.
The spectral radius of the infinite time step, which appears in the generalized-$\alpha$ time integration scheme, is set to 0.2 in this study.
We verified that this parameter has no discernible influence on the results reported in this study. 
The working of the generalized-$\alpha$ time-integration scheme \cite{jansen2000generalized} is explained in greater detail below.

First, we predict unknowns at the next time step as  $\dot{\bl u}_{n+1} = \frac{\gamma - 1}{\gamma} \dot{\bl u_n}$, $\bl u_{n+1} = \bl u_n$, and $p_{n+1} = p_n$. 
Next, the unknowns are computed at an intermediate time, namely $\dot{\bl u}_{n+\alpha_m}$, $\bl u_{n+\alpha_f}$ and $p_{n+\alpha_f}$ as
\begin{equation}
\begin{split}
   \dot{\bl{u}}_{n+\alpha_m}=\dot{\bl{u}}_{n}+\alpha_m(\dot{\bl{u}}_{n+1}-\dot{\bl{u}}_{n}), \\
   \bl{u}_{n+\alpha_f}=\bl{u}_{n}+\alpha_f(\bl{u}_{n+1}-\bl{u}_{n}), \\
   p_{n+\alpha_f}=p_{n}+\alpha_f(p_{n+1}-p_{n}), \\
   \end{split}
   \label{initiator}
\end{equation}
where $\alpha_f=(1+\rho_\infty)^{-1}$, $\alpha_m=(3-\rho_\infty)(2+2\rho_\infty)^{-1}$, and $\gamma=0.5+\alpha_m-\alpha_f$ are parameters that depend on the spectral radius of infinite time step $\rho_\infty$ (0.2 in our computations). 
These intermediate values are then employed to calculate $B_G$ and $B_S$ in Eq.~\eqref{rbvms_eq}.
After discretization and Newton-Raphson linearization, these equations yield
\begin{equation}
	 \begin{bmatrix}
	    \bl K & \bl G \\
	    \bl D & \bl L
	\end{bmatrix} \begin{bmatrix}
	    \bl \Delta \dot{\bl u} \\
	    \bl \Delta  p
	\end{bmatrix} 
	 = 
	- \begin{bmatrix}
	    \bl R_M \\
	    \bl R_C
	\end{bmatrix},
	\label{linear_sys_RBVMS}
\end{equation}
where,
\begin{equation}
\begin{split}
   \bl K & = \frac{\partial \bl R_M}{\partial \Delta \dot{\bl{u}}}, \\
   \bl G & = \frac{\partial \bl R_M}{\partial \Delta p}, \\
   \bl D &=  \frac{\partial \bl R_C}{\partial \Delta \dot{\bl{u}}}, \\
   \bl L & = \frac{\partial \bl R_C}{\partial \Delta p}, \\ 
   \end{split}
   \label{lin_eq_param}
\end{equation}
$\bl R_M$ and $\bl R_C$ are the momentum and continuity residuals and $\Delta \dot {\bl u}$ and $\Delta p$ are correction to our initial prediction of unknowns at the next time step.

The linear system in Eq. \eqref{linear_sys_RBVMS} is solved iteratively using the GMRES technique.
 Then corrections are applied using
\begin{equation}
\begin{split}
   \dot{\bl u}_{n+1}\leftarrow\dot{\bl u}_{n+1}+\Delta \dot{\bl u}, \\
   \bl u_{n+1}\leftarrow \bl u_{n+1}+\gamma \Delta t \Delta\dot{\bl u}, \\
   p_{n+1}\leftarrow p_{n+1}+\Delta p, \\
   \end{split}
   \label{corrector}
\end{equation}
Following these corrections, the Newton-Raphson iteration loop will be terminated if the norm of the residual is smaller than a specified tolerance. 
Otherwise, new intermediate variables are computed via Eq. \eqref{initiator} and a new iteration is performed.
As mentioned before, for the linear Stokes equation, only one iteration is enough. This entire process is repeated at each time step.

Overall, the RBVMS algorithm requires three nested iteration loops: the outer loop for time-stepping, the first inner loop for Newton-Raphson iterations, and the innermost iterative loop for the linear solver. 
For the problems investigated in this study, however, the nonlinear terms in Eq.~\eqref{rbvms_eq} can be neglected, resulting in two nested loops for the linear solver and time stepping.

Unlike operator splitting techniques that time advance velocity first and then enforce divergence-free condition by solving a Poisson equation, the RBVMS solves for both at the same time during a single linear solve as implied by Eq. \eqref{linear_sys_RBVMS}. 
Despite this difference, the sub-block matrix that relates pressure to continuity ($\bl L$ in Eq.~\eqref{linear_sys_RBVMS} that arises from $\bl u_{\rm p} \cdot \nabla q$ stabilization term in Eq. \eqref{rbvms_eq}) is identical to the discrete Laplacian operator that appears in the operator splitting technique. 
Note that this sub-block was zero in the linear system that arises from the SCVS formulation due to the lack of any stabilization term in our formulation (c.f., Eq. \eqref{linear_sys_def}).
As a consequence of this zero sub-block, the tangent matrix is ill-conditioned, resulting in a lower rate of convergence for the SCVS in comparison to the RBVMS. 

\section{MSS formulation} \label{app:mss}
 To allow for a more direct one-to-one comparison against the SCVS technique, we have implemented unsteady Stokes equations in the time domain using mixed shape functions. 
The weak formulation of this solver (called MSS) is identical to the SCVS except for the traditional treatment of the acceleration term. 
Namely, we use a conventional time integration scheme (i.e., the second-order implicit generalized-$\alpha$ method explained in~\ref{app:rbvms}) for this term. 
The weak form of the MSS states: find $\bl u \in \bl {\mathcal S}$ and $p \in \mathcal P$, such that for all $\bl w \in \bl {\mathcal W}$ and $q \in \mathcal Q$
\begin{equation}
   B_G\left(\bl w, q; \bl u, p\right) = F \left(\bl w, q\right), 
\end{equation}
where
\begin{equation}
    \begin{split}
    B_G & =  \int_\Omega \left[ \rho \bl w \cdot \dot {\bl u}  + \nabla \bl w : ( -p \bl I + \mu \nabla^{\rm s} \bl u ) + q \nabla \cdot \bl u \right] \rm d \Omega \\
   F & = \int_{\Gamma_{\rm h}} \bl w \cdot \bl h \rm d \Gamma. 
   \end{split}
   \label{mss_eq}
\end{equation}
Note that the test function and solution spaces are identical to those of the SCVS except for being defined in real rather than complex domain.  
Equation \eqref{mss_eq} is discretized using triangular and tetrahedral elements in 2D and 3D that were used for the SCVS.  
We use the same time integration scheme (i.e., generalized-$\alpha$) and linear solver as the RBVMS (i.e., GMRES).
The MSS also requires two nested loops: an outer loop for time-stepping, and an inner loop for the linear solver.
 \begin{table}[H]
 
     \caption{Comparison of the different solvers for the 3D pipe flow at $W=8\pi$ in terms of computational cost and overall error. 
     QM1 and LM1 meshes from Table \ref{tab:cylinder_msh} were used for these computations, with the former being used for the SCVS and MSS and the latter for the RBVMS. }
     \centering
     \begin{tabular}{cccc}
          \toprule
           & SCVS & MSS & RBVMS \\
          \midrule
          Mesh & QM1 & QM1 & LM1 \\
          The total computational cost (in CPU hours) & 1.25 & 1522 & 347\\
          The overall error; $e(T/2)$ & $6.3\times 10^{-3}$ & $1.79\times 10^{-2}$ & $4.1 \times 10^{-2}$\\
          \bottomrule
     \end{tabular}
     \label{tab:compare}
 \end{table}
  Table~\ref{tab:compare} compares the overall error (Eq. \eqref{total_error}) and the cost of MSS with the SCVS and RBVMS solvers for $W=8\pi$ and $t=T/2$. 
  The details of the meshes used for these computations are reported in Table \ref{tab:cylinder_msh}, where QM1 is employed for the SCVS and MSS and LM1 was adopted for the RBVMS. 
  The parameters of the time integration scheme are kept the same as the RBVMS, which were reported in Section \ref{sec:sol_procedure}.  
  
  The huge cost reduction from the MSS to SCVS can be explained by the fact that the SCVS solver performs just one linear solve, whereas the MSS solver performs 10,000 linear solves (each time step requiring one linear solve). 
  However, the performance gap between the MSS and SCVS is 1522/1.25 = 1,220 that is less than 10,000 as the average number of GMRES iterations for the MSS is less than that of the SCVS (i.e., 2,336.7 versus 7,943).
  Having pure real operations as oppose to complex operations in the GMRES algorithm also reduces the performance gap between the two. 
  
  To explain the lower performance of the MSS in comparison to the RBVMS, we must analyze the structure of the linear system for these two solvers. 
  Since there is no product of $p$ and $q$ in Eq. \eqref{mss_eq}, the discrete form of the continuity equation will not depend on the pressure. 
  This lack of dependence is in contrast to the RBVMS that through $\bl u_{\rm p} \cdot \nabla q$ term in Eq. \eqref{rbvms_eq} produces a non-zero $\bl L$ block in the tangent matrix in Eq. \eqref{lin_eq_param}. 
This non-zero block reduces the condition number of the tangent matrix, leading to a much faster convergence rate of the GMRES algorithm for the RBVMS (493.97 for the RBVMS versus 2,336.7 for the MSS). 

In terms of error, the SCVS that does not suffer from the time integration error and employs quadratic shape functions for velocity is the most accurate formulation.
The SCVS is followed by the MSS solution that contains time integration error and the RBVMS that on top of that error uses linear elements.

The comparison of the SCVS against the MSS is kept limited to this appendix. 
For the remainder of this article, the SCVS is only compared against the RBVMS for three primary reasons. 
Firstly, as confirmed by the results shown in Table \ref{tab:compare}, in general, the RBVMS performs better than the MSS, hence if the SCVS outperforms the RBVMS, it will also outperform the MSS. 
Secondly, the RBMVS is a more widely adopted formulation than the MSS, thus using it as a benchmark can provide a better point of reference. 
Thirdly, the cost and convergence properties of the MSS significantly suffer for larger meshes, making the simulation of some of the cases reported in Sections \ref{sec:cylinder_case} and \ref{sec:clinical_case} nearly impractical. 
For these reasons, we benchmark the SCVS against the RBVMS for all cases discussed under Section \ref{sec:results}.

\end{document}